\documentclass{IEEEtran}

\usepackage{amssymb}
\usepackage{stmaryrd}
\usepackage{amsmath}
\allowdisplaybreaks
\usepackage{amssymb}
\usepackage{graphicx}
\usepackage{todonotes}
\usepackage{cite}
\usepackage{algorithmic}
\usepackage{algorithm}
\usepackage{mathtools}
\usepackage{caption}
\usepackage{subcaption}
\usepackage{float}
\usepackage{comment}
\usepackage{makecell}
\usepackage{xcolor}
\usepackage{multirow}
\usepackage{bm}
\usepackage{bbm}
\usepackage{glossaries}
\usepackage{url}

\newcommand{\vect}[1]{\boldsymbol{\mathrm{#1}}}
\newcommand{\mat}[1]{\boldsymbol{\mathrm{#1}}}

\newcommand{\expt}[2]{\E_{#1}\left[#2\right]}
\newcommand{\cn}[2]{\ensuremath{\mathcal{C}\cN\left(#1,#2\right)}}

\newcommand{\maximize}{\mathop{\rm maximize}\limits}
\newcommand{\subjectto}{\mathop{\rm subject~to}\limits}
\newcommand{\minimize}{\mathop{\rm minimize}\limits}

\newcommand{\prob}[1][]{
\ifthenelse{\isempty{#1}}%
      {\ensuremath{P}}%
    {\ensuremath{P\left\(#1\right\)}}%
}

\def\mysubsubsection#1{\textbf{#1.}}
\def\defeq{\triangleq}

\newcommand{\transp}{{\sf T}}
\newcommand{\herm}{{\sf H}}

\def\ba{{\vect{a}}}

\def\bg{{\vect{g}}}
\def\bh{{\vect{h}}}

\def\bm{{\vect{m}}}

\def\bo{{\vect{o}}}

\def\bs{{\vect{s}}}

\def\bu{{\vect{u}}}
\def\bv{{\vect{v}}}
\def\bw{{\vect{w}}}
\def\bx{{\vect{x}}}
\def\by{{\vect{y}}}
\def\bz{{\vect{z}}}
\def\b0{{\vect{0}}}

\def\bzero{{\vect{0}}}

\def\bphi{{\vect{\phi}}}
\def\bpsi{{\vect{\psi}}}
\def\btheta{{\vect{\theta}}}

\def\bpi{{\vect{\pi}}}
\def\brho{{\vect{\rho}}}

\def\bD{{\mat{D}}}

\def\bG{{\mat{G}}}
\def\bH{{\mat{H}}}
\def\bI{{\mat{I}}}

\def\bQ{{\mat{Q}}}
\def\bR{{\mat{R}}}

\def\bV{{\mat{V}}}

\def\bX{{\mat{X}}}
\def\bY{{\mat{Y}}}

\def\bPsi{{\mat{\Psi}}}

\def\bGamma{\mat{\Gamma}}

\def\bPi{\mat{\Pi}}

\def\cA{{\mathcal{A}}}

\def\cK{{\mathcal{K}}}
\def\cL{{\mathcal{L}}}

\def\cN{{\mathcal{N}}}

\def\cS{{\mathcal{S}}}
\def\cT{{\mathcal{T}}}
\def\cU{{\mathcal{U}}}

\def\cX{{\mathcal{X}}}

\def\E{{\mathbb{E}}}

\def\C{{\mathbb{C}}}
\def\ind{{\mathbbm{1}}}

\newtheorem{propi}{Proposition}
\newtheorem{remark}{Remark}

\newtheorem{assumption}{Assumption}
\newtheorem{definition}{Definition}
\newenvironment{proof outline}{\paragraph*{Proof Outline}}{\hfill$\IEEEQEDopen$}

\def\pilotlength{L}
\def\act{\textup{act}}
\def\p{\textup{p}}
\def\d{\textup{d}}

\def\SINR{\mathsf{SINR}}
\def\effSINR{\overline{\mathsf{SINR}}}

\def\th{\textup{th}}
\def\suc{\textup{suc}}
\def\pilotspace{\cA_1\times\cdots\times\cA_N}
\def\activeusers{\overline{\cU}}
\def\expectedrate{\overline{R}}
\def\BibTeX{{\rm B\kern-.05em{\sc i\kern-.025em b}\kern-.08em
    T\kern-.1667em\lower.7ex\hbox{E}\kern-.125emX}}
\AtBeginDocument{\definecolor{ojcolor}{cmyk}{0.93,0.59,0.15,0.02}}

\newacronym{gfra}{GFRA}{grant-free random access}
\newacronym{mimo}{MIMO}{multiple-input multiple-output}
\newacronym{bs}{BS}{base station}
\newacronym{marl}{MARL}{multi-agent reinforcement learning}
\newacronym{urllc}{URLLC}{ultra-reliable low-latency communication}
\newacronym{rl}{RL}{reinforcement learning}
\newacronym{lsfc}{LSFC}{large-scale fading coefficient}
\newacronym{mmse}{MMSE}{minimum mean-square error}
\newacronym{sinr}{SINR}{signal-to-noise-plus-interference ratio}
\newacronym{mr}{MR}{maximum ratio}
\newacronym{zf}{ZF}{zero-forcing}
\newacronym{cdf}{CDF}{cumulative density function}
\newacronym{csma}{CSMA}{carrier-sense multiple access}
\newacronym{noma}{NOMA}{non-orthogonal multiple access}
\newacronym{ncpdr}{NCPDR}{normalized cumulative packet drop rate}
\newacronym{cbl}{CBL}{contextual bandit learning}

\makeglossaries

\begin{document}

\title{Delay-Constrained Grant-Free Random Access in MIMO Systems: Distributed Pilot Allocation and Power Control}

\author{Jianan~Bai, Zheng~Chen, and~Erik~G.~Larsson
\thanks{This paper was presented in part at the Asilomar SSC 2021 conference~\cite{conf}.}
\thanks{
The authors are with the Department of Electrical Engineering (ISY), Link\"oping University, 58183 Link\"oping, Sweden (email: jianan.bai@liu.se, zheng.chen@liu.se, erik.g.larsson@liu.se). This work was supported in part by Excellence Center at Link\"oping-Lund in Information Technology (ELLIIT), and by the Knut and Alice Wallenberg (KAW) foundation. The computations were enabled by resources provided by the National Academic Infrastructure for Supercomputing in Sweden (NAISS) and the Swedish National Infrastructure for Computing (SNIC) partially funded by the Swedish Research Council through grant agreements no. 2022-06725 and no. 2018-05973.
}}

\maketitle

\begin{abstract}
	We study a delay-constrained grant-free random access system with a multi-antenna base station. The users randomly generate data packets with expiration deadlines, which are then transmitted from data queues on a first-in first-out basis.
    To deliver a packet, a user needs to succeed in both random access phase (sending a pilot without collision) and data transmission phase (achieving a required data rate with imperfect channel information) before the packet expires.
    We develop a distributed, cross-layer policy that allows the users to dynamically and independently choose their pilots and transmit powers to achieve a high effective sum throughput with fairness consideration.
    Our policy design involves three key components:
    1) a proxy of the instantaneous data rate that depends only on macroscopic environment variables and transmission decisions, considering pilot collisions and imperfect channel estimation; 
    2) a quantitative, instantaneous measure of fairness within each communication round; 
    and 3) a deep learning-based, multi-agent control framework with centralized training and distributed execution.
    The proposed framework benefits from an accurate, differentiable objective function for training, thereby achieving a higher sample efficiency compared with a conventional application of model-free, multi-agent reinforcement learning algorithms.
    The performance of the proposed approach is verified by simulations under highly dynamic and heterogeneous scenarios.

\end{abstract}

\begin{IEEEkeywords}
	Grant-free random access, delay constraint, MIMO, fairness, and distributed control.
\end{IEEEkeywords}

\section{Introduction}

\Gls{urllc} is anticipated to facilitate a variety of emergent applications such as remote surgery and autonomous vehicles~\cite{sachs20185g}. 
Conventional grant-based scheduling fails to meet the delay requirements due to its excessive handshake overhead, often surpassing the tolerable 1-millisecond delay.
\Gls{gfra} is a promising solution to reduce uplink latency~\cite{3gpp.36.300}. 
In \gls{gfra}, users can transmit payload data together with metadata (pilot and other signaling) without waiting for permission or scheduling information.
Despite the advantages of \gls{gfra}, a major challenge is the allocation of pilot sequences to users, and the handling of pilot collisions during the uplink access, which inevitably results if there are more users than available orthogonal pilots.
Additionally, the transmit power needs to be properly selected to ensure that the data packets can be successfully delivered with minimal inter-user interference.

In this paper, we consider the problem of pilot selection and power control in a \gls{mimo}-enabled \gls{gfra} system. The problem is complicated by the need for a cross-layer modeling and the uncoordinated nature of \gls{gfra}. 
We aim to develop a distributed policy such that users can dynamically and independently select their pilots and transmit powers by using only local information to maximize the network performance and provide fairness among users.
We propose to solve this problem using deep learning, which can learn a complicated policy without relying on a usually restrictive model \cite{she2021tutorial}.
Different learning paradigms can be applied in different scenarios -- supervised learning for approximating known policies with labeled data; unsupervised learning for cases where an explicit objective function can be obtained~\cite{sun2020unsupervised}; \gls{rl} for making sequential decisions when neither labeled data nor an explicit objective function is available.

Among various learning paradigms, \gls{marl} appears to be the most relevant, and it has been successfully applied to develop distributed policies in wireless networks (e.g., \cite{9204466,8815511,8986647,deng2022reinforcement,le2022tinyqmix}).
However, conventional \gls{marl} schemes were developed for general-purpose tasks and may not provide the most efficient solution to our particular use case. 
To be specific, they suffer from: 
i) delayed and sparse rewards (the immediate reward might not accurately evaluate actions in the long run); 
ii) incapability of satisfying instantaneous constraints; 
iii) the multi-agent credit assignment problem (a global reward may not reflect an individual contribution); 
and iv) a high demand for samples (an accurate sample-based estimation is required for the expected return, which is difficult to obtain for large search spaces).

Model-based learning has demonstrated effectiveness across various applications \cite{wang2019benchmarking}, and one could expect further performance improvements by integrating specific domain knowledge into the algorithm design. 
As we will see shortly, for our problem, we possess strong domain knowledge: 
i) the collision probability using a stochastic pilot selection policy can be calculated; 
ii) the success probability of payload transmission for a given power allocation can be well approximated; and
iii) the stochastic optimization problem can be (approximately) solved by solving a sub-problem in each decision stage with an objective function that more precisely evaluates the actions.

\subsection{Related Work}
When using mutually orthogonal pilots, several approaches to pilot allocation and collision resolution for random access have been proposed.  
For example, the possibility of using multiple or superimposed pilots, to effectively retain the pilot orthogonality, was investigated in \cite{8770292,8069007,choi2020approach}. 
To improve the collision resolution, another line of work (e.g., \cite{bai2020collision}) exploited channel hardening and favorable propagation properties of massive \gls{mimo} and used successive interference cancellation to recover the collided signals. 
Strategies that assign users unique but mutually non-orthogonal pilots were investigated in, for example,~\cite{sparse-sp-gf} along with associated collision resolution algorithms based on compressed sensing techniques. 
A comparative analysis of the use of orthogonal versus non-orthogonal pilots was presented in~\cite{ding2019analysis}. 
The results suggest that the performance of non-orthogonal pilots, which reduces pilot collision at the expense of degraded channel estimation quality compared to the case of orthogonal pilots, is contingent on the specific scenario. 
Specifically, non-orthogonal pilots may underperform when requiring high data rates.
Studying non-orthogonal pilots is not the main focus of our paper, but we will provide some numerical comparisons as a baseline.
Non-coherent transmission schemes and unsourced communication systems (e.g.,~\cite{9374476,fengler2022pilot}) are beyond our scope.

Applying \gls{marl} in \gls{gfra} systems has received increasing attention.  
A pilot selection policy was developed in \cite{9204466} with significant improvements in the average aggregate throughput compared with various baseline schemes. 
However, \cite{9204466} considered only a non-dynamic system without delay constraints and data rate requirements. 
In \cite{8815511}, a \gls{csma} system with a single channel was considered, wherein each user selects its access probability based on the urgency of their packets and system load. 
A transmission tax was introduced to decouple the multi-agent training for improved scalability.
A clustering-based sub-channel selection and (discrete) power control policy was designed in \cite{8986647} for a \gls{noma} system to maximize the long-term throughput.
In \cite{deng2022reinforcement}, the authors considered the coexistence of ALOHA users and users that employ a learned random access policy with delay-constrained traffic. 
A distributed policy for dynamic resource selection is developed in \cite{le2022tinyqmix} for a lightly loaded system with a relatively large delay tolerance.
To the best of our knowledge, there has not been a research work that considers joint pilot selection and (continuous) power control for a realistically modeled \gls{mimo}-assisted \gls{gfra} system with stringent delay requirements.
Additionally, most research in this direction applies conventional model-free \gls{marl} algorithms without efficiently exploiting the model knowledge to accelerate the learning process.

\vspace{-0.2cm}
\subsection{Contributions and Organization of the Paper}

\noindent \textbf{1) Cross-Layer Modeling:}

We present the physical layer and the network layer models of the system in Section \ref{sec: system}. 
Particularly, in the physical layer, we characterize the instantaneous data rate of users, for both \gls{mr} and \gls{zf} receive combining, with a \gls{mmse} channel estimator and pilot collisions.
To eliminate the dependence of the rate expression on the random small-scale fading for policy design, we develop a rate proxy that depends only on the macroscopic environment variables and the transmission decisions of users.
To the best of our knowledge, the rate proxy for \gls{zf} under pilot collisions is new.

\noindent \textbf{2) Quantification of Fairness:}

We study min-max fairness of the system by minimizing the (normalized) packet drop rate of the worst performing user in Section \ref{sec: formulation}.
The original formulation of the problem is a stochastic network optimization problem, which involves the time average of the stochastic packet drop processes with time dependence imposed by the evolution of data queues that cannot be fully predicted.
To overcome this challenge, we develop two approximations to the problem that can be solved immediately in each decision stage. Additionally, we reveal a unified structure behind these two approximations, and interpret it as a sum-priority maximization.
Specifically, the priority level of each user takes accounts of both its previous access results and the current queue status.
The (normalized) sum-priority provides an accurate, instantaneous quantification of fairness within each communication round.

\noindent \textbf{3) Deep Learning-Based Distributed Policy Design:}

To exactly maximize the sum-priority, the users still need to share information (e.g., priority levels) to each other or to a central server, which contradicts the open-loop operations of \gls{gfra}.
Therefore, we propose a deep learning-based distributed control framework that requires centralized training but enables distributed execution in Section \ref{sec: policy design}.
This learning framework is motivated by \gls{marl}, while significantly deviating from conventional \gls{marl} by employing an unsupervised training scheme.
Particularly, by exploiting our results above, we obtain a learning objective (the expected sum-priority) that is directly differentiable with respect to the policy parameters, which obviates the need for a sample-based estimate of the expected reward over the joint action space.
This objective function also accurately measures individual contributions so that the credit assignment problem is naturally alleviated. 
The framework learns a hybrid policy that combines (discrete) pilot selection and (continuous) power control.

\textbf{Remark:}
Part of this work was presented in the conference paper  \cite{conf}, where we considered only the pilot transmission in a simplified collision model and assumed that packet delivery is successful whenever the pilot transmission is successful. In this paper, we consider a much more realistic scenario with data rate requirements and incorporate power control.

\subsection{Notation}
Vectors are denoted by boldface lowercase letters, $\bx$, matrices by boldface uppercase letters, $\bX$, and sets by calligraphic letters, $\cX$, with cardinality $|\cX|$. The superscripts $(\cdot)^\transp$, $(\cdot)^\herm$, $(\cdot)^*$, and $(\cdot)^{-1}$ denote transpose, conjugate transpose, complex conjugate, and inverse, respectively. 
$\E[\cdot]$ denotes the statistical expectation. $\ind\{\cdot\}$ is the indicator function, which equals to $1$ for true propositions and $0$ otherwise. $\C^n$ denotes the space of $n$-dimensional complex vectors. The multivariate circularly symmetric complex Gaussian distribution with covariance matrix $\bR$ is denoted by $\cn{\bzero}{\bR}$. $\bD_{\bx}$ denotes a diagonal matrix with $\bx$ on its diagonal. 
$\|\cdot\|$ denotes the Euclidean vector norm.

\section{System Model}
\label{sec: system}

We consider the uplink of a single-cell narrowband wireless system. 
The \gls{bs} has $M$ receive antennas and serves $N$ machine-type devices (users) located within its coverage area. 
Time is divided into equal-length slots. 
We adopt the block-fading assumption, i.e., the channels remain constant during a slot (consisting of $\tau$ symbols) and vary independently across different slots.\footnote{We choose the block-fading model for simplicity and tractability. More realistic channel models that consider intra-block variations or inter-block correlation (for example, those in \cite{truong2013effects,3gpp.36.901}) are left for future work.}
The uplink data of each user are divided into equal-size packets and the transmission duration of each packet is one slot. 
We model the packet arrivals at each user by a Bernoulli process, i.e., a new packet is generated at user $i$ with probability $\lambda_i$ in each slot.
Each user, $i\in\cN\defeq\{1,\cdots,N \}$, has a data queue to store the generated data packets, with the queue backlog in slot $t$ denoted by $Q_{it}$. 
We define the set of backlogged users (those with non-empty queues) by $\cK_t\defeq\{i: Q_{it}>0 \}$. 
Each backlogged user, $i\in\cK_t$, can decide whether to access at the beginning of the slot.

There are $L$ mutually orthogonal pilots $\bphi_1,\cdots,\bphi_L\in\C^L$, each normalized to have unit energy such that $\|\bphi_l\|=1$ for all $l\in\cL\defeq\{1,\cdots,L\}$. 
We require $\pilotlength < N$ due to the limited channel coherence so that the users cannot be pre-assigned unique, mutually orthogonal pilots. 
Pilot collision occurs when multiple users select the same pilot. 
The pilot selection of each backlogged user $i\in\cK_t$ is represented by $a_{it}\in \{0\}\cup\cL$, where $a_{it}=0$ denotes the decision to back off, and $a_{it}=l\in\cL$ indicates that the $l$-th pilot is selected. 
For an idle user $i\notin\cK_t$, we set $a_{it}=0$ by default. Additionally, we define $\cU_{lt}\defeq\{i:a_{it}=l\}$ as the set of users that select the $l$-th pilot, and $\activeusers_t\defeq \cU_{1t}\cup\cdots\cup\cU_{Lt}$ as the set of active users (those who transmit any of the pilots).

\subsection{Physical Layer Model}

\subsubsection{Pilot Detection}
During pilot transmission in slot $t$, the received pilot signal, $\by_{mt}^\p\in\C^{\pilotlength}$, at the $m$-th antenna is
\begin{equation}
\label{eq: received pilot}
\begin{aligned}
	\by_{mt}^\p &= \sum_{i\in\activeusers_t} \sqrt{\pilotlength\beta_i\rho_i^\p} h_{imt} \bphi_{a_{it}} + \bw_{mt}^\p\\
	&= \sum_{l\in\cL}\sum_{i\in\cU_{lt}} \sqrt{\pilotlength\beta_i\rho_i^\p} h_{imt} \bphi_{l} + \bw_{mt}^\p,
\end{aligned}
\end{equation}
where $\beta_i$ represents the \gls{lsfc} of user $i$ (similar to \cite{marzetta2016fundamentals}, $\beta_i$ is normalized such that the noise has unit variance), $h_{imt}\sim\cn{0}{1}$ represents the small-scale fading coefficient that is assumed to be independent across users and antennas, $\rho_i^\p\in[0,\rho_{\max}]$ is the transmit power of the pilot signal, and $\bw_{mt}^\p\sim\cn{\bzero}{\bI_L}$ is additive noise that is independent across antennas.

We consider channel inversion power control for pilot transmission, i.e.,
$\rho_i^\p = ({\beta_{\min} }/{\beta_i})\rho_{\max}$,
where $\beta_{\min}\defeq\min_{i\in\cN}\{\beta_i\}$. 
This gives $\pilotlength\beta_i\rho_i^\p=\pilotlength\beta_{\min}\rho_{\max}\defeq \rho_0$. We further define the effective channel coefficient of pilot $l$ as
\begin{equation}
	g_{lmt}\defeq \frac{1}{\sqrt{|\cU_{lt}|}}\sum_{i\in\cU_{lt}}h_{imt} \sim \cn{0}{1}
\end{equation}
when $|\cU_{lt}|\geq 1$, and define $\{g_{lmt}\}$ as independent $\cn{0}{1}$ random variables for the case when $|\cU_{lt}|=0$. Notice that $g_{lmt}=h_{imt}$ when $\cU_{lt}=\{i\}$, which holds for all non-collided users. We can then re-write \eqref{eq: received pilot} as
\begin{equation}
\label{eq: received pilot in vector form}
	\by_{mt}^p = \sum_{l\in\cL} \sqrt{\rho_0|\cU_{lt}|} g_{lmt} \bphi_l + \bw_{mt}^\p.
\end{equation}

For activity detection (the process of identifying the active users by processing the received pilot signals), due to the orthogonality of pilots, we de-spread the received signal by
\begin{equation}
	\bphi_l^\herm\by_{mt}^\p = \sqrt{\rho_0|\cU_{lt}|}g_{lmt} + \bphi_l^\herm\bw_{mt}^\p,
\end{equation}
where $\bphi_l^\herm\bw_{mt}^\p\sim\cn{0}{1}$ since the pilots have unit energy.
$\bphi_l^\herm\by_{mt}^\p$ has distribution $\cn{0}{\rho_0|\cU_{lt}| + 1}$ and is independent across different antennas. Therefore, we have
\begin{equation}
	\frac{1}{M}\sum_{m=1}^M |\bphi_l^\herm\by_{mt}^\p|^2 \xrightarrow{M\rightarrow\infty} \rho_0|\cU_{lt}| + 1
\end{equation}
by the law of large numbers. 
When the number of antennas is sufficiently large so that the channel hardens, the multiplicity of the transmitted pilots, i.e., $\bu_t \defeq [|\cU_{1t}|,\cdots,|\cU_{Lt}|]^\transp$, can be accurately determined by energy detection~\cite{ding2021dynamic}. 
Since activity detection is not the main focus of our paper, and to simplify the analysis, we assume perfect pilot detection. 

\begin{assumption}
\label{as: pilot}
The multiplicities of the transmitted pilots, $\bu_t$, is known. When a pilot is transmitted by exactly one user, i.e., $|\cU_{lt} |=1$, the identity of that user can be known.
\end{assumption}

\subsubsection{Channel Estimation}
Define the set of active pilots as $\cL_t^\act\defeq\{l:|\cU_{lt}|\geq 1\}$.
Since we cannot identify the collided users, we choose to estimate the effective channel coefficients $\{g_{lmt}\}$ for each active pilot, instead of estimating the actual channel coefficients $\{h_{imt}\}$ for each active user.
Notice that this makes no difference for non-collided users. 
The \gls{mmse} estimate of $g_{lmt}$ is given by
\begin{equation}
\label{eq: channel estimate orthogonal}
	\widehat{g}_{lmt} = \frac{\sqrt{\rho_0|\cU_{lt}|}}{\rho_0|\cU_{lt}| + 1}\bphi_l^\herm\by_{mt}^\p,
\end{equation}
and the mean-square of the channel estimate is 
\begin{equation}
\label{eq: channel estimation covariance for O-GFRA}
	c_{lt} \defeq \E[|\widehat{g}_{lmt}|^2] = \frac{\rho_0|\cU_{lt}|}{\rho_0|\cU_{lt}| + 1}.
\end{equation}
By the orthogonality principle, the channel estimation error $\widetilde{g}_{lmt}\defeq g_{lmt}-\widehat{g}_{lmt}$ is uncorrelated (and, therefore, independent under Rayleigh fading) with $g_{lmt}$. 
Also, the mean-square estimation error is given by $1 - c_{lt}$.

\subsubsection{Payload Data Transmission}

During the data transmission phase, the received signal at the \gls{bs} is given by
\begin{equation}
\begin{aligned}
    \by_t =& \sum_{i\in\activeusers_t}\sqrt{\beta_i\rho_{it}}q_{it}\mathbf{h}_{it} + \bw_t,
\end{aligned}
\end{equation}
where $\rho_{it}\in[0,\rho_{\max}]$ represents the transmit power (notice that we assumed channel inversion power control only for the pilot transmission), $\bh_{it}\defeq[h_{i1t},\cdots,h_{iMt}]^\transp$ is the channel vector, $q_{it}$ is the transmitted data symbol with unit energy which is uncorrelated across users, and $\bw_t\sim\cn{\bzero}{\bI}$ is the noise vector.

We denote by $\bg_{lt}\defeq[g_{l1t},\cdots,g_{lMt}]^\transp$ the effective channel of the $l$-th pilot over all antennas.
Analogously, we define $\widehat{\bg}_{lt}$ and $\widetilde{\bg}_{lt}$ as the estimate and estimation error of $\bg_{lt}$.
For a non-collided user $i$, that satisfies $\bh_{it} = \bg_{a_{it}t} = \widehat{\bg}_{a_{it}t} + \widetilde{\bg}_{a_{it}t}$, we perform receive combining by using a combining vector $\bv_{it}$ to obtain (the collided users are not interesting since they cannot be identified)
\begin{equation}
\begin{aligned}
	\bv_{it}^\herm\by_t =& \underbrace{\sqrt{\beta_i\rho_{it}}q_{it}\bv_{it}^\herm\widehat{\bg}_{a_{it}t}}_{\text{desired signal}} + \sqrt{\beta_i\rho_{it}}q_{it}\bv_{it}^\herm\widetilde{\bg}_{a_{it}t}\\
	&+ \sum_{j\in\activeusers_t\backslash i}\sqrt{\beta_j\rho_{it}}q_{it}\bv_{it}^\herm\mathbf{h}_{jt} + \bv_{it}^\herm\bw_t.
\end{aligned}
\end{equation}
The instantaneous \gls{sinr} of that user is given by
\begin{equation}
\label{eq: instantaneous sinr}
	\SINR_{it} =
	\frac{\beta_i\rho_{it}|\bv_{it}^\herm\widehat{\bg}_{a_{it}t}|^2}{\beta_i\rho_{it}|\bv_{it}^\herm\widetilde{\bg}_{a_{it}t}|^2 + \sum_{j\in\activeusers_t\backslash i}\beta_j\rho_{jt}|\bv_{it}^\herm\mathbf{h}_{jt}|^2 + \|\bv_{it} \|^2}.
\end{equation}

For ease of notation, we define a superscript $(\cdot)^\act$.
\begin{definition}
\label{def: act}
	For a matrix $\bX$, or a vector $\bx$, with at least one dimension corresponding to the pilot indices $\cL$, we define a reduced-dimensional matrix $\bX^\act$, or a vector $\bx^\act$, by keeping only the entries corresponding to active pilots $\cL_t^\act$ in $\bX$ or $\bx$. 
	Conversely, when $\bX^\act$ or $\bx^\act$ is defined first, $\bX$ or $\bx$ represents the extended matrix or vector by filling the missing entries corresponding to the inactive pilots with zeros. 
\end{definition}

We consider both \gls{mr} and \gls{zf} combining, by introducing the combining matrix
\begin{equation}
	\bV_t^\act \defeq \left\{
    \begin{array}{ll}
        \widehat{\bG}_t^\act, & \text{MR} \\
        \widehat{\bG}_t^\act \left((\widehat{\bG}_t^\act)^\herm\widehat{\bG}_t^\act \right)^{-1} , & \text{ZF}
    \end{array}
    \right.,
\end{equation}
where $\widehat{\bG}_t\defeq[\widehat{\bg}_{1t},\cdots,\widehat{\bg}_{Lt}]$, and taking the $a_{it}$-th column as the combining vector $\bv_{it}$, i.e., $\bv_{it} \defeq [\bV_t]_{:,a_{it}}$. 

For a targeted decoding error probability, we approximate the instantaneous achievable data rate of user $i$ by
\begin{equation}
\label{eq: instantaneous rate}
	R_{it} = \log_2(1+\ell \cdot \SINR_{it}),
\end{equation}
where $\ell\in(0,1]$ is a penalty factor accounting for the effects of finite blocklength\footnote{A more accurate characterization of the finite-blocklength effect can be obtained using, for example, the normal approximation in \cite[Th. 55]{polyanskiy2010channel}. Since an accurate finite-blocklength analysis is not our focus, we use the approximation in \eqref{eq: instantaneous rate}. However, our approach can be applied as long as the rate expression is a non-increasing, convex function of the ${1}/{\SINR}$.} and the coding and modulation scheme.
Such an approximation has been used in, for example, \cite{qiu1999performance}, and $1/\ell$ is also known as the \gls{sinr} gap \cite{garcia2006snr}.

Recall that each user has fixed-size packets corresponding to a fixed instantaneous rate requirement.
Denoting the rate threshold of user $i$ as $R_i^\th$, we make the following assumption.
\begin{assumption}
\label{as: data}
	A non-collided user $i$ can successfully deliver its head-of-line packet if $R_{it} \geq R_i^\th$. 
\end{assumption}

Finally, we define the success indicator of user $i$, based on Assumptions \ref{as: pilot} and \ref{as: data}, as
\begin{equation}
\label{eq: success indicator}
	\mu_{it} \defeq \ind\{|\cU_{a_{it}t}|=1\}\cdot\ind\{R_{it} \geq R_i^\th\}.
\end{equation}

\subsubsection{Rate Proxy for Algorithm Training}

The instantaneous rate expression in \eqref{eq: instantaneous rate} depends on the random small-scale channel fluctuations, which cannot be acquired by the users when making transmission decisions. Instead, we look for a rate metric that depends only on the macroscopic environment variables and transmission decisions (e.g., \glspl{lsfc}, pilot selection, and power control),
and will be using $\E[R_{it}]$ (more precisely, its lower bounds for tractability) as a proxy for $R_{it}$, where the expectation is taken over all small-scale channel fluctuations. (Notice that we will always use the instantaneous rate in \eqref{eq: instantaneous rate} for simulations. The expressions developed here are used only for algorithm design.)

By noticing that $\log(1+1/x)$ is a convex function, we can apply the Jensen's inequality to obtain
\begin{equation}
\label{eq: expected rate lower bound}
	\E[R_{it}] \geq \overline{R}_{it} \defeq \log(1 + \ell \cdot \effSINR_{it} )
\end{equation}
where 
\begin{equation}
	\effSINR_{it} \defeq \left(\expt{}{\frac{1}{\SINR_{it}}} \right)^{-1}.
\end{equation}

\begin{propi}
\label{propi: effective SINR}
\begin{equation}
\label{eq: effective SINR}
	\effSINR_{it} = 
	\left\{\begin{array}{ll}
		 \displaystyle\frac{(M-1)c_{a_{it}t}\beta_i\rho_{it}}{\sum_{j\in\activeusers_t}\beta_j\rho_{jt} - c_{a_{it}t}\beta_i\rho_{it} + 1}, & \text{MR}\\
		 \displaystyle\frac{(M-|\cL_t^\act|)c_{a_{it}t}\beta_i\rho_{jt}}{ \sum_{j\in\activeusers_t}\big(1-\frac{c_{a_{it}t}}{|\cU_{a_{jt}t}|} \big)\beta_j\rho_{jt} + 1},& \text{ZF}
	\end{array}
	\right..
\end{equation}
\end{propi}

\begin{IEEEproof}
	The result for \gls{mr} follows immediately from \cite[Appendix D]{marzetta2016fundamentals}. The result for \gls{zf} is proved in the Appendix.
\end{IEEEproof}

For $\overline{R}_{it}$ to be an accurate approximation to $R_{it}$, the instantaneous \gls{sinr} in \eqref{eq: instantaneous sinr} should be sufficiently concentrated around $\effSINR_{it}$. 
Unfortunately, we might not always have enough concentration. 
To see this, examine the numerator and the denominator in \eqref{eq: instantaneous sinr} separately with a normalized $\bv$, i.e., $\|\bv\|=1$. 
In the numerator, the random variable $|\bv_{it}^\herm\widehat{\bg}_{a_{it}t}|^2/M$ concentrates for both \gls{mr} and \gls{zf} as $M\rightarrow \infty$.
However, in the denominator, the terms $|\bv_{it}^\herm\widetilde{\bg}_{a_{it}t}|^2$ and $\{|\bv_{it}^\herm\bh_{jt}|^2\}_{i\neq j}$ might not concentrate.
Take \gls{mr} combining for example, $|\bv_{it}^\herm\widetilde{\bg}_{a_{it}t}|^2$ and $\{|\bv_{it}^\herm\bh_{jt}|^2\}_{i\neq j}$ become independent exponential random variables. 
Unless $|\activeusers_t|$ is sufficiently large, the denominator does not necessarily concentrate.
The problem can be alleviated for \gls{zf}, when good channel estimates are obtained so that the interference can be considerably suppressed.
But a general conclusion is that, when making short packet transmissions, one may not be able to benefit from a concentrated \gls{sinr} even in massive \gls{mimo}.
Fortunately, as we will observe in the numerical results, approximating $\SINR_{it}$ by $\effSINR_{it}$ still results in a useful algorithm.

\subsection{Network Layer Model}

Recall that we consider the random packet arrivals at each user $i\in\cN$, modeled as a Bernoulli process with rate $\lambda_i$. 
Once generated, the packets are backlogged in the queue of that user. 
Additionally, to account for the timeliness of data packets, we assume that every packet of user $i$ is associated with a maximum tolerable delay (also referred to as a deadline), denoted as $d_i^{\max}$, that is defined as the number of time slots within which a newly generated packet has to be delivered to the destination before expiration. 
For simplicity, we assume that all packets of user $i$ have the same maximum tolerable delay so that each queue operates in a first-in-first-out manner. 
We define $d_{it}\in\{1,\cdots,d_i^{\max}\}$, for all $i\in\cK_t$, to be the number of remaining time slots (including the current one) of the head-of-line packet at slot $t$ before it expires. 
When the queue is empty, i.e., $i\notin\cK_t$, we set $d_{it}=0$ by default. 
A packet is discarded if it cannot be successfully delivered before the deadline. 
We therefore define the packet drop indicator as
\begin{equation}
	D_{it} \defeq (1-\mu_{it})\ind\{d_{it}=1\}.
\end{equation}

Let $\{\gamma_{it}\}$ denote the packet arrival process, where each $\gamma_{it}$ is modeled as a Bernoulli random variable with $\expt{}{\gamma_{it}}=\lambda_i$ and is independent across users and slots.  
Also, define the packet departure process $\{b_{it}\}$ given by 
\begin{equation}
	b_{it} \defeq \mu_{it} + D_{it},
\end{equation}
which equals 1 when $\mu_{it}=1$ or $D_{it}=1$, and 0 otherwise.
The evolution of the queue backlog of user $i$ is described by
\begin{equation}
	Q_{i,t+1} = \max\{Q_{it}-b_{it},0\} + \gamma_{it}.
\end{equation}

\section{Min-Max Fairness}
\label{sec: formulation}

We consider a fairness perspective of the system, formulated as a stochastic network optimization problem that minimizes the (normalized) packet drop rate of the worst-performing user. 
To obviate the difficulties in directly solving this problem, we propose two approximations, one using a log-sum-exp approximation of the max function, and the other employing the Lyapunov drift-plus-penalty framework. 
These two approaches give a unified, quantitative measure of instantaneous fairness, interpreted as the ``sum-priority'' of the successful users.

\subsection{Stochastic Formulation}

We define the effective throughput of user $i$ as the average number of data packets it successfully delivers per time slot, $\lambda_i-\overline{D}_i$, where $\overline{D}_i$ is the packet drop rate defined as
\begin{equation}
    \overline{D}_i \defeq \limsup_{T\rightarrow\infty} \E\left[\frac{1}{T}\sum_{t=1}^T D_{it} \right].
\end{equation}
Here, the expectation is taken over the randomness of the packet arrival process $\{\gamma_{it}\}$, and the packet departure process $\{b_{it}\}$. 
To maximize the effective throughput of a user, we can equivalently minimize the packet drop rate.

Each user is associated with a drop rate threshold $D_i^\th$, which represents the quality of service (QoS) requirement. 
We then formulate the stochastic min-max fairness problem as\footnote{In addition to the fractional objective ${\overline{D}_i}/{D_i^\th}$, the proposed approach also works for other objectives, e.g., ${\overline{D}_i} - {D_i^\th}$.}
\begin{equation}
\tag{P}
\label{P}
\begin{aligned}
    &	\minimize_{\{\ba_t,\brho_t\}}&&\max_{i\in\cN}\left\{\frac{\overline{D}_i}{D_i^\th} \right\}   \\
    &	\subjectto\!\! &&(\ba_t,\brho_t)\in\cA\times[0,\rho_{\max}]^N\\
    & &&\forall t\in\{1,2,\cdots \},
\end{aligned}
\end{equation}
where $\ba_t\defeq[a_{1t},\cdots,a_{Nt}]^\transp$ and $\brho_t\defeq[\rho_{1t},\cdots,\rho_{Nt}]^\transp$ are joint pilot and power allocations in slot $t$, and $\cA \defeq \pilotspace$, with $\cA_i=\{0\}\cup\cL$. 
Notice that $\overline{D}_i$ is a stochastic function of all joint decisions $\{\ba_t\}$ and $\{\brho_t\}$ across time.

It is infeasible to directly solve \eqref{P} to obtain an optimal sequential decision solution due to the following two reasons:
\begin{itemize}
    \item The problem \eqref{P} involves the time average of the stochastic processes $\{D_{it}\}$ with time dependence imposed by the evolution of data queues, which cannot be fully predicted.
    \item The max function in \eqref{P} requires the determination of worst-performing user $i^* = \arg\max_{i}\{\overline{D}_i / D_i^\th\}$ for all feasible decisions, which is a combinatorial problem.
\end{itemize}

In what follows, we develop two approaches to solve \eqref{P} approximately by constructing a time-varying objective (that combines both the previous access results and the urgency of undelivered packets) and \emph{greedily} optimizing the objective in every slot to make real-time decisions that depend only on the current state of the system.

\begin{remark}
    By ``greedy'', we mean making decisions based on only local or immediate information, without considering the impact on future time instances \cite[pp. 64]{sutton2018reinforcement}.
    It can greatly reduce the complexity of a real-time decision-making process.
    Meanwhile, a greedy approach can still perform well, even in the long run, if the immediate objective is properly chosen. 
    For example, in Q-learning, \emph{greedily} selecting actions to maximize the Q-function (if accurately estimated) is optimal in the long run, as the Q-function represents the long-term return.

\end{remark}

\subsection{The First Approach: Log-Sum-Exp}

We approximate the max function in \eqref{P} by the log-sum-exp function as in \cite{chen2013markov}, i.e.,
\begin{equation}
	\max_{i\in\cN}\{x_i\} \approx \frac{1}{\alpha}\log \left(\sum_{i\in\cN} \exp(\alpha x_i) \right),
\end{equation}
where $\alpha\in(0,\infty)$ can be interpreted as an ``inverse temperature''. As shown in \cite[pp. 72]{boyd2004convex}, the approximation gap is upper-bounded by $\frac{1}{\alpha}\log N$, and the approximation becomes an exact equality if $\alpha\rightarrow\infty$.

By applying the log-sum-exp approximation and limiting our focus to a finite frame $\cT\defeq\{1,\cdots,T\}$ with $T$ slots, we obtain the following  problem
\begin{equation}
\begin{aligned}
    &\minimize_{\{\ba_t,\brho_t\}}~&& \frac{1}{\alpha}\log\left(\sum_{i\in \cN} \exp\left(\frac{\alpha}{TD_i^{\th}}\sum_{t=1}^{T}D_{it} \right)\right)\\
    &\subjectto && (\ba_t,\brho_t)\in\cA\times[0,\rho_{\max}]^N,~~\forall t\in\mathcal{T}.
    \label{P: frame}
\end{aligned}
\end{equation}

To simplify \eqref{P: frame}, we remove the logarithm (the problem will not change due to the monotonicity of the logarithm) and define the \gls{ncpdr}
\begin{equation}
    \xi_{it}=\frac{1}{TD_i^{\th}}\sum_{t'=1}^{t}D_{it'},\quad\forall i\in\cN,
\end{equation}
where $TD_i^\th$ can be interpreted as the total ``budget'' of packet drops of user $i$ in a frame, and $\xi_{it}$ is the ratio of currently consumed budget till slot $t$. Then, \eqref{P: frame} can be re-written as
\begin{equation}
\begin{aligned}
    &\minimize_{\{\ba_t,\brho_t\}}~&& \sum_{i\in \cN} f\left(\xi_{iT} \right)\\
    &\subjectto && (\ba_t,\brho_t)\in\cA\times[0,\rho_{\max}]^N,~~\forall t\in\mathcal{T},
    \label{P: frame2}
\end{aligned}
\end{equation}
where we introduce a fairness-promoting function
\begin{equation} \label{eq: fairness func}
	f(x) \defeq \frac{\exp{(\alpha x)} - 1}{\exp(\alpha) - 1},
\end{equation}
such that $\alpha\rightarrow\infty$ leads to strict min-max fairness, $\alpha\rightarrow 0$ gives $f(x)=x$ and the problem becomes a sum-drop-rate minimization, and $\alpha\in(0,\infty)$ gives an elastic level of fairness among different devices. 
(The function  $f(x)$ is a normalized version of $\exp(\alpha x)$. 
This does not change the problem but permits a better interpretation as $\alpha\rightarrow0$.)
Problem \eqref{P: frame2} admits a straightforward interpretation -- the cost associated with a user is determined by its final \gls{ncpdr} through a mapping defined by the fairness-promoting function.

Obtaining an optimal sequence of decisions requires one to know all the packet arrivals and channel conditions in the frame, which is still infeasible.
We obviate this difficulty and make real-time decisions by greedily solving the following problem in each slot
\begin{equation}
\begin{aligned}
    &\minimize_{\ba_t,\brho_t}~&& \sum_{i\in \cN} f\left(\xi_{i,t-1} + \frac{1}{TD_i^\mathrm{th}}D_{it}(\ba_t,\brho_t) \right)\\
    &\subjectto && (\ba_t,\brho_t)\in\cA\times[0,\rho_{\max}]^N,
    \label{P: frame3}
\end{aligned}
\end{equation}
where we write $D_{it}$ as $D_{it}(\ba_t,\brho_t)$ to accentuate that it is an explicit function of the joint decision $(\ba_t,\brho_t)$. 
Similar notations will be used henceforth.

The formulation in \eqref{P: frame3} ignores the prior information about future (potential) packet drops, which has already been included in the expiration time $d_{it}$ of the head-of-line packet. 
(One may also consider the expiration time of other packets in the queue and the arrival rates.) 
Therefore, similar to \cite{fountoulakis2020dynamic}, we replace $D_{it}(\ba_t,\brho_t)$ by $\widetilde{\delta}_{it}(1-\mu_{it}(\mathbf{a}_t,\brho_t))$, where
\begin{equation}
    \widetilde{\delta}_{it} \defeq 1 - \frac{d_{it}-1}{d_i^{\max}}.
    \label{eq: urgency level}
\end{equation}
Here, $\widetilde{\delta}_{it}$ can be interpreted as an urgency level to deliver the head-of-line packet in the queue. 
Notice that $\widetilde{\delta}_{it}(1-\mu_{it}(\mathbf{a}_t,\brho_t))$ and $D_{it}(\mathbf{a}_t,\brho_t)$ take the same extreme values
\begin{equation*}
    \left\{
    \begin{array}{ll}
        0:& \text{ if } \mu_{it}(\mathbf{a}_t,\brho_t)=1\\
        1:& \text{ if } \mu_{it}(\mathbf{a}_t,\brho_t)=0 \text{ and } d_{it}=1,
    \end{array}
    \right. 
\end{equation*}
while the former can be seen as a softened version of the latter one by assigning non-zero values in between to incorporate the prior information of future packet drops.

Finally, we approximate \eqref{P: frame} by using a series of sub-problems that will be solved in each slot $t\in\mathcal{T}$: 
\begin{equation}
\begin{aligned}
    &\maximize_{\ba_t,\brho_t}~~&&\sum_{i\in\cK_t} \widetilde{\eta}_{it}^{(\textup{S1})}\mu_{it}(\mathbf{a}_t,\brho_t)\\
    &\subjectto && (\ba_t,\brho_t)\in\cA\times[0,\rho_{\max}]^N,
    \label{S1}
\end{aligned}
\tag{S1}
\end{equation}
where $\widetilde{\eta}_{it}^{(\textup{S1})}$ is defined by\footnote{Notice that $f\big(x+y\mu\big)=f(x)+\big(f(x+y)-f(x)\big)\mu $ for $\mu\in\{0,1\}$.}
\begin{equation}
\label{eq: eta1}
    \widetilde{\eta}_{it}^{(\textup{S1})} \defeq f\big(\xi_{i,t-1} + {\delta}_{it} \big) - f\left(\xi_{i,t-1} \right)
\end{equation}
with the normalized urgency level ${\delta}_{it}\defeq\widetilde{\delta}_{it}/(TD_i^\th)$.

\subsection{The Second Approach: Virtual-Queue}

By introducing an auxiliary variable, $\overline{z}>0$, Problem \eqref{P} can be expressed in epigraph form as
\begin{equation}
\begin{aligned}
    &	\minimize_{\{\ba_t,\brho_t\},\overline{z}}~&&\overline{z}  \\
    &	\subjectto && \overline{D}_i \leq \overline{z} D_i^\th,~\forall i\in\cN\\
    & &&(\ba_t,\brho_t)\in\cA\times[0,\rho_{\max}]^N\\
    & &&\forall t\in\{1,2,\cdots \}.
\end{aligned}
\label{P2}
\end{equation}
We introduce a bounded stochastic process $z_t\in[0,z_{\max}]$, such that $\limsup_{T\rightarrow\infty}\E[\frac{1}{T}\sum_{t=1}^T z_t]=\overline{z}$. 
Then, problem \eqref{P2} can be transformed into
\begin{subequations}
\begin{align}
    \minimize_{\{\ba_t,\brho_t,z_t\}}~~~ & \limsup_{T\rightarrow\infty}\E\left[\frac{1}{T}\sum_{t=1}^T z_t\right]\label{subeq: obj}  \\
    \subjectto~~ & \limsup_{T\rightarrow\infty}\E\left[\frac{1}{T}\sum_{t=1}^T\left(\frac{D_{it}}{D_i^{\th}}-z_t \right)\right]\leq 0 \label{subeq: cons}\\
    &(\ba_t,\brho_t)\in\cA\times[0,\rho_{\max}]^N,~\forall t\in\{1,2,\cdots \}\nonumber\\
    &0 \leq z_t \leq z_{\max},~\forall t\in\{1,2,\cdots \}.\nonumber
\end{align}
\end{subequations}
The constraint \eqref{subeq: cons} can be transformed into a queue stability problem. 
To see this, we assign each user a virtual queue. 
The vector of virtual queue backlogs (one should distinguish this from the data queue backlog $Q_{it}$) of all users is denoted as $\bX_t\defeq[X_{1t},\cdots,X_{Nt}]^\transp$, where the virtual queue backlog of user $i$ is updated by
\begin{equation}
\label{eq: virtual queue evolution}
    X_{i,t+1} = \max\left\{X_{it}-z_t,0 \right\} + \frac{D_{it}}{D_i^\th}.
\end{equation}
The constraint in \eqref{subeq: cons} is satisfied if $X_{it}$ is rate stable~\cite{neely2010stochastic}, i.e., 
    $\lim_{t\rightarrow\infty}{X_{it}}/{t}=0$ almost surely, for all $i\in\cN$.

Denote by $\bGamma_t=\big(\bX_t,\vect{d}_t\big)$ the network state, where $\vect{d}_t=[d_{1t},\cdots,d_{Nt}]^\transp$ contains the packet deadlines. To establish queue stability, we consider the conditional Lyapunov drift 
\begin{equation}
    \Delta_t \defeq \E\big[\varphi\big(\bGamma_{t+1} \big) - \varphi\big(\bGamma_t \big)|\bGamma_t\big],
    \label{eq: Lyapunov drift}
\end{equation}
where $\varphi\big(\bGamma_t \big)\defeq\frac{1}{2}\sum_{i\in\cN}X_{it}^2$ is a quadratic Lyapunov function. We further consider the drift-plus-penalty function
\begin{equation}
    \Delta_t + V\E\big[z_t|\bGamma_t \big],
    \label{eq: DPP}
\end{equation}
where $V>0$ is a factor controlling the trade-off between the queue stability and the optimality of the objective in \eqref{subeq: obj}.
The drift-plus-penalty \eqref{eq: DPP} is upper bounded by
\begin{align}
\label{eq: bound on DPP}
    \Delta(t)& + V\E\big[z_t|\bGamma_t \big] \nonumber \\ 
    =& \frac{1}{2}\E\left[\sum_{i\in\cN}\left(X_{i,t+1}^2 - X_{it}^2\right) \Big| \mat{\Gamma}_t \right] + V\E\big[z_t|\bGamma_t \big] \nonumber  \\ 
    \leq& {\frac{1}{2}\E\left[\sum_{i\in\cN}\left(z_t^2+\left(\frac{D_{it}}{D_i^\textup{th}}\right)^2 \right)\Big| \mat{\Gamma}_t \right]} \nonumber \\
    &+ \E\left[\sum_{i\in\cN}X_{it}\left(\frac{D_{it}}{D_i^\textup{th}} - z_t \right)\Big| \mat{\Gamma}_t \right] + V\E\big[z_t|\bGamma_t \big] \nonumber  \\
    \leq& \underbrace{\frac{N}{2}z_{\max}^2 + \frac{N^2}{2}\sum_{i\in\cN}\left(\frac{1}{D_i^\textup{th}} \right)^2}_{\text{constant}} \nonumber\\
    &+ \E\!\left[\sum_{i\in\cN}\frac{X_{it}}{D_i^\th}D_{it} + \left(V\!-\! \sum_{i\in\cN}X_{it}\right)\!z_t \Big| \bGamma_t \right]. 
\end{align}

We approximate Problem \eqref{P2} by greedily minimizing the upper bound of the drift-plus-penalty function in \eqref{eq: bound on DPP}. This leads to a sequence of subproblems in each slot $t\in\cT$:
\begin{equation}
\begin{aligned}
    &\minimize_{\ba_t,\brho_t,z_t} &&\underbrace{\sum_{i\in\cN}\frac{X_{it}}{D_i^\th}D_{it}(\mathbf{a}_t,\brho_t)}_{\text{depends only on }(\ba_t,\brho_t)} + \underbrace{\left(V\!-\! \sum_{i\in\cN}X_{it}\right)\!z_t}_{\text{depends only on }z_t} \label{subeq: obj2}  \\
    &\subjectto && (\ba_t,\brho_t)\in\cA\times[0,\rho_{\max}]^N\\
    & && 0 \leq z_t \leq z_{\max}.
\end{aligned}
\end{equation}
The first term in the objective function of \eqref{subeq: obj2} depends only on $(\ba_t,\brho_t)$, and the second term depends only on $z_t$. 
Thus, we can solve for the optimal $(\ba_t,\brho_t)$ and for the optimal $z_t$ separately. Minimizing the second part gives
\begin{equation}
\label{eq: z}
    z_t = z_{\max} \cdot \ind\left\{\sum\nolimits_{i\in\cN}X_{it}> V \right\},
\end{equation}  
which will be used in \eqref{eq: virtual queue evolution} for updating the virtual queue backlog $X_{it}$.

Similar to problem \eqref{S1}, we replace $D_{it}(\ba_t,\brho_t)$ by $\widetilde{\delta}_{it}(1-\mu_{it}(\ba_t,\brho_t))$ in the first term, where $\widetilde{\delta}_{it}$ is defined in \eqref{eq: urgency level}, to incorporate the prior information on future packet drops. 
This gives the following problem
\begin{equation}
\begin{aligned}
    &\maximize_{\ba_t,\brho_t}~~&&\sum_{i\in\cK_t} \widetilde{\eta}_{it}^{(\textup{S2})}\mu_{it}(\mathbf{a}_t,\brho_t),\\
    &\subjectto && (\ba_t,\brho_t)\in\cA\times[0,\rho_{\max}]^N,
    \label{S2}
\end{aligned}
\tag{S2}
\end{equation}
where $\widetilde{\eta}_{it}^{(\textup{S2})}$ is calculated by
\begin{equation}
\label{eq: eta2}
    \widetilde{\eta}_{it}^{(\textup{S2})} = {X_{it}\delta_{it}}.
\end{equation}
Notice that the virtual queue backlog $X_{it}$ in \eqref{eq: eta2} plays a similar role as $\xi_{it}$ in \eqref{eq: eta1} -- they both incorporate historical information about previous access results so that a user with larger $X_{it}$ or $\xi_{it}$ will be prioritized. 
However, it is less straightforward to interpret how the parameters $V$ and $z_{\max}$ affect the evolution of $X_{it}$.
Roughly speaking, from \eqref{eq: virtual queue evolution} and \eqref{eq: z}, we observe that $V$ determines how frequently $X_{it}$ is updated, and $z_{\max}$ determines how significant each update can be (their effects can not be separated though). By choosing a small $V$ and a large $z_{\max}$,  the history will be discarded rapidly. Conversely, a large $V$ with a small $z_{\max}$ keeps a long history.

\subsection{A Unified Perspective: Sum-Priority Maximization}
\label{subsec: sub-priority}

The approximated problems \eqref{S1} and \eqref{S2} share a unified structure, where the coefficient $\widetilde{\eta}_{it}^{(s)}$, for $s\in\{\textup{S1},\textup{S2}\}$, can be interpreted as the priority level of user $i$ in slot $t$. 
A solution $(\vect{a}_t,\vect{\rho}_t)$ is mapped to the success indicators $\{\mu_{it} \}$ in \eqref{eq: success indicator}. 
An optimal solution should maximize the sum-priority of the successful users. 
These two approximation approaches differ in how the priority levels are defined:

\subsubsection*{Log-Sum-Exp}
\begin{figure}[t]
    \centering
    \includegraphics[width=7cm]{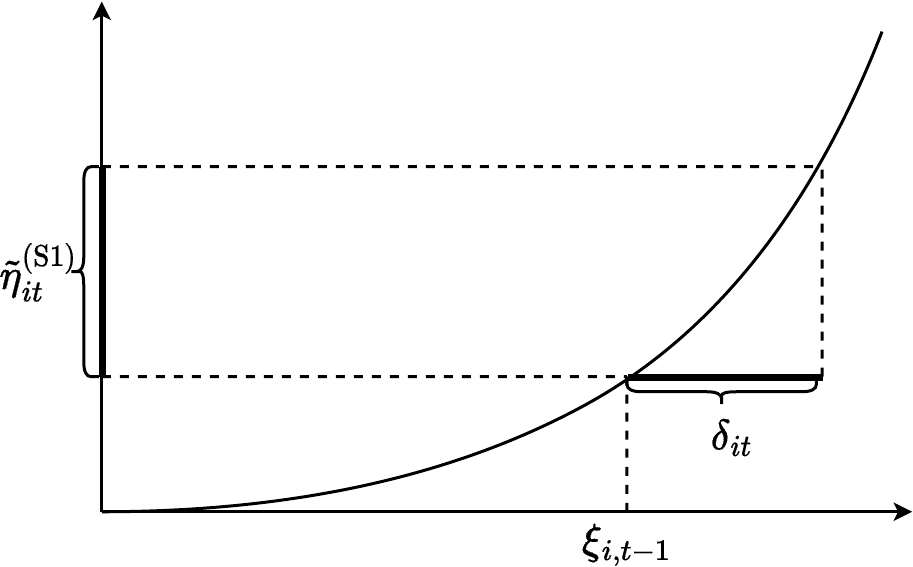}
    \caption{An illustration of the mapping defined by the fairness promoting function in \eqref{S1}.}
    \label{fig:fairness function}
\end{figure}

In \eqref{S1}, we introduce the fairness-promoting function, $f(\cdot)$, to provide a mapping from the NCPDR, $\xi_{i,t-1}$, and the normalized urgency level, ${\delta}_{it}$, to the priority level $\widetilde{\eta}_{it}$. 
See the illustration in Fig.~\ref{fig:fairness function}. 
One can observe that as $\xi_{i,t-1}$ increases, $\widetilde{\eta}_{it}$ grow more rapidly with ${\delta}_{it}$. 
The impact of $\xi_{i,t-1}$ is controlled by the inverse temperature $\alpha$. 
As $\alpha\rightarrow0$, the impact of $\xi_{i,t-1}$ disappears, and we obtain a sum-drop-rate minimization problem.

\subsubsection*{Virtual-Queue}
The interpretation of \eqref{S2} becomes more straightforward in the extreme case when $V\rightarrow\infty$. 
The virtual queue backlog is now given by $X_{it} = \sum_{t'=1}^{t-1}D_{it'}/{D_i^\textup{th}}=T\xi_{i,t-1}$. 
The priority level becomes $\widetilde{\eta}_i(t) = T\xi_{i,t-1}{\delta}_{it}$. 
One can see that \eqref{S2} also defines a mapping from $\xi_{i,t-1}$ and $\widetilde{\delta}_{it}$ to the priority level $\widetilde{\eta}_{it}$, and the same argument holds: 
a larger $\xi_{i,t-1}$ makes $\widetilde{\eta}_{it}$ grow more rapidly with $\widetilde{\delta}_{it}$.

\begin{remark}
    Exactly solving \eqref{S1} and \eqref{S2} still requires the users to share information to each other or to a central server, contradicting the open-loop nature of \gls{gfra}.
    We circumvent this need by developing a deep learning framework to enable the users to learn a distributed access policy (through centralized, offline training) that approximates the solution to \eqref{S1} and \eqref{S2} by using only their local information in Section~\ref{sec: policy design}.
\end{remark}

\section{Distributed Policy Design}
\label{sec: policy design}

We now shift our focus to the development of a policy for joint pilot selection and power control in the GFRA system introduced in Section \ref{sec: system}. 
By a ``policy'', we mean a mapping from situations to decisions, which can be either deterministic, stochastic, or mixed.
Notice that the approximations, \eqref{S1} and \eqref{S2}, developed in Section \ref{sec: formulation} essentially define two different policies. 
That is, by knowing the \emph{global} information, denoted $\bs$, that consists of the priority levels, the queue status, and the LSFCs of all users, solving \eqref{S1} or \eqref{S2} gives a global control decision $(\ba,\brho)$. 
Nevertheless, a global policy that requires knowing $\bs$ cannot be implemented for GFRA, since the users only have access to their local information and, potentially, some limited feedback information. 
We therefore look for a distributed policy where each user $i$ uses its local information $\bo_{i}$ to generate its own control decision $(a_{i},\rho_{i})$.

\mysubsubsection{Notation}
We will reuse variables defined in Sections \ref{sec: system} and \ref{sec: formulation}, but with slight changes. 
First, since time-dependent information of the environment is encapsulated within the global state $\bs$, we will omit the time index in the subscript when considering only a single slot. 
When multiple slots are considered, we will write the global state in slot $t$ as $\bs_t$ and the decisions as $(\ba_t,\brho_t)$. 
Second, we will explicitly write out the variables' dependence on $\ba$, $\brho$, and $\bs$. 
For example, we write the success indicator $\mu_{it}$ as $\mu_i(\ba,\brho|\bs)$, and the priority level $\widetilde{\eta}_{it}$ as $\widetilde{\eta}_i(\bs)$.

\mysubsubsection{Assumptions} 
In this section, we will introduce several assumptions when characterizing the (approximate) transmission success probability. These assumptions will be used only for algorithm design. The simulation environment will be fully based on the system model presented in Section \ref{sec: system}.

\subsection{Expected Sum-Priority}

We consider a stochastic pilot selection policy, where user $i$ chooses $a_i=l$ with probability $\pi_{il}$, and $\sum_{l=0}^L\pi_{il}=1$. 
The matrix of pilot selection probabilities is denoted by $\bPi \defeq [\bpi_1,\cdots,\bpi_N]$, where $\bpi_{i}\defeq[\pi_{i0},\pi_{i1},\cdots,\pi_{iL}]^\transp$. 
Under a global state $\bs$, we aim to obtain a joint policy $(\bPi,\brho)$ to maximize the expected (normalized) sum-priority 
\begin{equation}
\label{eq: expected sum-priority}
	J(\bPi,\brho|\bs) \defeq \sum_{i\in\cN} \eta_i(\bs) P_i^\suc(\bPi,\brho|\bs),
\end{equation}
where $\eta_i(\bs) \defeq {\widetilde{\eta}_i(\bs)}/{\sum_{j\in\cN}\widetilde{\eta}_j(\bs)}$ is the normalized priority level of user $i$, and $P_i^\suc(\bPi,\brho|\bs) = \expt{}{\mu_i(\ba,\brho|\bs)}$ is the success probability with the expectation taken by randomly sampling $\ba$ using the probabilities in $\bPi$ and by averaging over small-scale channel fluctuations. 
Based on the definition in \eqref{eq: success indicator}, the success probability can be calculated by
\begin{equation}
\begin{aligned}
	P_i^\suc(\bPi,\brho|\bs) = &\underbrace{\Pr\left\{|\cU_{a_i}(\ba)|=1 \right\}}_{\defeq P_i^\p(\bPi)}\\&\cdot\underbrace{\Pr\left\{R_i(\ba,\brho|\bs) \geq R_i^\th\big||\cU_{a_i}(\ba)|=1\right\}}_{\defeq P_i^\d(\bPi,\brho|\bs)},
\end{aligned}
\end{equation}
where $P_i^\p(\bPi)$ is the probability that user $i$ transmits a pilot without collision, i.e., $|\cU_{a_i}(\ba)|=1$, and $P_i^\d(\bPi,\brho|\bs)$ is the probability that the instantaneous data rate requirement is satisfied, i.e., $R_i(\ba,\brho|\bs)\geq R_i^\th$, when user $i$ is non-collided.
The non-collision probability is given by
\begin{equation}
	P_i^\p(\bPi) = \sum_{l\in\cL}\pi_{il} \prod_{j\in\cN\backslash i} (1-\pi_{jl}).
\end{equation}

Now we proceed to characterize the probability of successful data transmission $P_i^\d(\bPi,\brho|\bs)$. 
Recall that the instantaneous data rate $R_i(\ba,\brho|\bs)$ of a non-collided user $i$ is given by \eqref{eq: instantaneous rate}. 
The characterization of $P_i^\d(\bPi,\brho|\bs)$ requires us to take two sources of randomness into account: the random pilot selection decisions according to the probabilities in $\bPi$, and the small-scale channel fluctuations. 
It appears infeasible to obtain a tractable expression for $P_i^\d(\bPi,\brho|\bs)$. 
Therefore, we approximate the instantaneous achievable data rate by the rate proxy in \eqref{eq: expected rate lower bound}. 

Since $\log_2(1+\ell x)$ is an increasing function of $x$ for $\ell>0$, the rate condition $\expectedrate_i(\ba,\brho|\bs) \geq R_i^\th$ is equivalent to 
\begin{equation}
\label{eq: SINR threshold}
	\frac{1}{\effSINR_i(\ba,\brho|\bs)} \leq \omega_i \defeq \frac{\ell}{2^{R_i^\th}-1}.
\end{equation}
By substituting \eqref{eq: effective SINR} into \eqref{eq: SINR threshold} and by defining the coefficients $\{\sigma_{ji}(\ba)\}$ in Table \ref{tab: power coefficients}, we obtain
\begin{equation}
\label{eq: interference power constraint}
	\sum_{j\in\cN\backslash i} \ind\{a_j\neq 0\}\sigma_{ji}(\ba)\beta_j\rho_j + 1 \leq \sigma_{ii}(\ba)\beta_i\rho_i,
\end{equation}
where we can interpret the LHS as interference-plus-noise power that scales with the transmit power of the interfering users, and the RHS as the interference tolerance of user $i$ that scales with its transmit power. 
Additionally, the coefficients $\{\sigma_{ji}(\ba)\}_{j\neq i}$ control how fast the interference power grows with $\{\rho_j\}_{j\neq i}$, and $\sigma_{ii}(\ba)$ determines how large $\rho_i$ is needed to overpower the interference. 
One can observe that $\{\sigma_{ji}(\ba)\}$ have a very complicated dependence on the pilot selection decision $\ba$. We avoid this dependence by making additional approximations. 

\begin{table*}[t]
	\centering
    \scalebox{1.2}{
	\begin{tabular}{|c|c|c|}
		\hline
		& \textbf{MR} & \textbf{ZF} \\
		\hline
		$\sigma_{ji}(\ba),~j\neq i$ & $1$ & $\displaystyle 1- \frac{\rho_0(\bs)}{\rho_0(\bs)|\cU_{a_i}(\ba)| + 1}$ \\
		\hline
		$\sigma_{ii}(\ba)$ & $\displaystyle  \frac{(M-1)\omega_i  \rho_0(\bs)|\cU_{a_i}(\ba)| - 1}{\rho_0(\bs)|\cU_{a_i}(\ba)| + 1} $ & $\displaystyle  \frac{(M-|\cL^\act(\ba)|)\omega_i \rho_0(\bs)|\cU_{a_i}(\ba)| - 1}{\rho_0(\bs)|\cU_{a_i}(\ba)| + 1} $  \\
		\hline
		$\varsigma_{ji}(\bs),~j\neq i$ & $1$ & $\displaystyle 1-\frac{\rho_0(\bs)}{\rho_0(\bs) + 1}$\\
		\hline
		$\varsigma_{ii}(\bs)$ & $\displaystyle\frac{(M-1)\omega_i \rho_0(\bs) - 1}{\rho_0(\bs) + 1} $ & $\displaystyle \frac{(M-L)\omega_i \rho_0(\bs) - 1}{\rho_0(\bs) + 1} $ \\
		\hline
	\end{tabular}
    }
	\caption{The expressions of $\{\sigma_{ji}(\ba)\}$ and $\{\varsigma_{ji}(\bs)\}$. }
	\label{tab: power coefficients}
\end{table*}

As shown in \eqref{eq: channel estimation covariance for O-GFRA} and Table \ref{tab: power coefficients},
$\{\sigma_{ji}(\ba)\}$ depend on $\ba$ through $\{|\cU_{a_i}(\ba)|\}$ for MR, and, additionally, on $|\cL^\act(\ba)|$ for ZF. 
We postulate that a good control policy should efficiently utilize all available pilots, i.e., $|\cL^\act(\ba)|\approx L$, without any pilot collisions, i.e., $|\cU_l(\ba)|\approx 1$ for all $l\in\cL$. 
This gives $c_{a_{it}}(\ba)\approx \rho_0(\bs)/(1+\rho_0(\bs))$. 
Notice that we might underestimate the impact of the pilot collisions.
By making these approximations, we replace $\{\sigma_{ji}(\ba)\}$ by $\{\varsigma_{ji}(\bs)\}$  in Table \ref{tab: power coefficients}.

Notice that by taking the stochastic pilot selection policy, $\ind\{a_j\neq 0\}$ is a Bernoulli random variable which is equal to one with probability $1-\pi_{i0}$. The LHS in \eqref{eq: interference power constraint} is a weighted sum of independent Bernoulli random variables with unequal non-zero probabilities, whose closed-form \gls{cdf} is generally very complicated~\cite{tang2019poisson}. To obtain a more tractable expression, we use the normal approximation, where the LHS in \eqref{eq: interference power constraint} can be approximated as a normal random variable with mean
\begin{equation}
	\mathsf{E}(\bPi,\brho|\bs) \defeq \sum_{j\in\cN\backslash i} \varsigma_{ji}(\bs)\beta_j\rho_j(1-\pi_{i0}) + 1,
\end{equation}
and variance
\begin{equation}
	\mathsf{Var}(\bPi,\brho|\bs) \defeq \sum_{j\in\cN\backslash i} \varsigma_{ji}^2(\bs)\beta_j^2\rho_j^2\pi_{i0}(1-\pi_{i0}).
\end{equation}
The probability of successful data transmission is then approximated as 
\begin{equation}
	P_i^\suc(\bPi,\brho|\bs) = 1 - S\left(\frac{\varsigma_{ii}(\bs)\beta_i\rho_i - \mathsf{E}(\bPi,\brho|\bs)}{\sqrt{\mathsf{Var}(\bPi,\brho|\bs)}} \right),
\end{equation}
where $S(\cdot)$ is the complementary \gls{cdf} of the standard normal distribution.

\subsection{Learning-Based Distributed Policy Optimization}

We have obtained a closed-form approximation to the expected sum-priority in \eqref{eq: expected sum-priority}. 
However, since the objective function does not decouple across users, the optimal decision of each user depends on the decisions of other users. 
It is still intractable to find the optimal distributed policy that maximizes the expected sum priority by using only the users' local information. 
We therefore consider a deep learning-based approach to this problem. 
Each user $i$ has a deep neural network (referred to as a policy network) with parameter $\btheta_i$ as the policy generator. 
Once an observation $\bo_{it}$ (the local information) is received in slot $t$, it is fed into the policy network to generate the outputs $\left(\bpi_{\btheta_i}(\bo_{it}),\rho_{\btheta_i}(\bo_{it})\right)$.
The observation in slot $t$ is 
\begin{equation}
	\bo_{it} = [d_{it}, \gamma_{it}, \beta_{it}, \nu_{it}]^\transp,
\end{equation}
where $d_{it}$ is the expiration time of the head-of-line packet, $\gamma_{it}$ is the packet arrival indicator, $\beta_{it}$ is the (normalized) LSFC in dB, and $\nu_{it}$ is set to the NCPDR $\xi_{i,t-1}$ for the log-sum-exp approximation in \eqref{S1} and the virtual queue backlog $X_{it}$ for \eqref{S2}. 
A feedback information $\bm_t$ broadcast by the \gls{bs} in each slot can also be included in the input to the policy network. 
An example of the feedback information can be found in \cite{9204466}, consisting of a ternary indicator (idle, collision, and successful transmission) for each pilot in the previous slot.

\begin{figure}[t]
    \centering
    \includegraphics[width=0.9\linewidth]{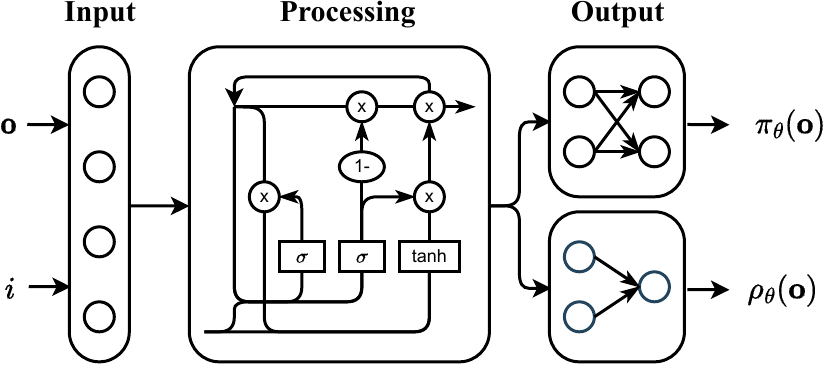}
    \caption{The structure of the policy network.}
    \label{fig: neural net}
\end{figure}

The policy network consists of an input module, a processing module, and an output module. 
The input module is a feedforward layer with ReLU activation. The processing module is a gated recurrent unit (GRU) layer to address the partial observability of agents~\cite{hausknecht2015deep}. 
The output module has two sub-modules for generating the pilot selection probabilities and the transmit power, respectively. Each sub-module consists of two feedforward layers with ReLU activation in the first layer. 
The outputs from the pilot selection sub-module have dimension $L+1$ and are normalized by the Softmax function to produce $\vect{\pi}_i(\vect{o}_i)$.
The last layer of the power allocation sub-module has a single neuron with Sigmoid activation and the output is scaled by $\rho_{\max}$ to generate the transmit power. 
The pilot selection action is randomly sampled using the generated probabilities. The neural network is sketched in Fig.~\ref{fig: neural net}.

We denote the neural network parameters of all devices by $\mat{\Theta} = \{\vect{\theta}_i \}$. Since the joint policy is a function of $\mat{\Theta}$ and the network state $\vect{s}$, we re-write the expected sum-priority in \eqref{eq: expected sum-priority} as $J(\mat{\Theta}|\vect{s})$.
The policy networks are jointly trained in an unsupervised manner to maximize the expected sum-priority over all possible network states. To incorporate the temporal correlation, we consider training using sequences of state transitions, and the training objective becomes
\begin{equation}
\label{eq: training obj with expectation}
	\maximize_{\mat{\Theta}}~~ \E_{\{\vect{s}_t\}_{t\in\mathcal{T}}}\left[ \sum_{t\in\mathcal{T}} J(\mat{\Theta}|\vect{s}_t) \right].
\end{equation}
To obtain an estimate of the expectation in \eqref{eq: training obj with expectation}, we collect the generated state transitions in a replay buffer during each training epoch. We run a fixed number of training iterations using a stochastic gradient descent (SGD)-based optimizer by sampling a mini-batch, $\mathcal{S}$, of state transitions:
\begin{equation}
\label{eq: train obj}
	\maximize_{\mat{\Theta}}~~ \frac{1}{|\mathcal{S}|} \sum_{\{\vect{s}_t \}\in \mathcal{S}}\sum_{t\in\mathcal{T}} J(\mat{\Theta}|\vect{s}_t) .
\end{equation}
Centralized, offline training is performed to update the parameters $\mat{\Theta}$ in an unsupervised manner.

To accelerate the training process, we use {parameter sharing}, such that all users share the same policy network, i.e., $\vect{\theta}_i=\vect{\theta}$ for all $i\in\cN$. 
To distinguish different agents and keep a more accurate history of the dynamics of the environment, the input to the policy network also contains the one-hot encoded agent index and the action selected in the last time slot.

During execution, a device only needs to feed its observation into the trained model in each slot to make the transmission decision. 
The execution is efficient and does not incur significant delays, as the neural network is lightweight with a short inference time.
The execution is also fully distributed, and no interaction is needed between users.

\subsubsection*{Pilot Pre-Allocation}
One critical issue of learning in multi-agent systems is that the global state and action spaces grow exponentially with the number of agents. 
This ``curse of dimensionality'' can make the problem exceedingly challenging or even intractable. 
One remedy is to limit the interactions between different agents. In \cite{qu2020scalable}, for example, a networked system was considered, where the agents are associated with a graph and interact only with their connected agents in the graph. 
In our \gls{gfra} system, the interactions can be limited by pre-allocating a subset of pilots to a group of users and letting different groups use disjoint subsets of pilots so that users from different groups will never collide.

\subsection{Relation to RL}
Our proposed learning approach is related to RL in terms of learning ``a mapping from situation to actions \cite{sutton2018reinforcement}''  through the interaction between agents and environment. 
However, there are some key differences.
In RL, the agents receive a ``reward'' from the environment after taking an action. The reward is a non-differentiable scalar that does not reflect the long-term effects of the actions.
The goal of RL is to maximize the cumulative reward over time, which requires a sample-based estimation of a value function that represents the expected sum of future rewards.
The estimation of the value function requires exploration by taking random actions and becomes challenging when the state and action spaces are large, as in our case.
In contrast, our approach has a differentiable training objective, i.e., the expected sum-priority, which is an explicit function of the policy and also reflects the long-term effects of the actions to some extent (through incorporating the urgency levels of packets).
By directly maximizing the expected sum-priority for the generated state sequences in an unsupervised manner, we avoid the exploration problem and the sample-based estimation of the value function, thereby achieving a higher sample efficiency.
We provide a numerical comparison between our approach and RL in Section \ref{sec: comparison with RL}.

\section{Simulations}

We evaluate the proposed approach in a single-cell system, where the (hexagonal) cell radius is $1$~km, and the \gls{bs} has $M=100$ receive antennas. The devices are dropped uniformly at random in the cell with a circular exclusion zone around the \gls{bs} of radius 0.05~km. For each device, the actual (unnormalized) LSFC is generated by
$
    \widetilde{\beta}_{i}=-140.6 - 36.7\log_{10}(\textup{dist}_i) +  \Upsilon_{i}
$
in dB, where $\textup{dist}_i$ is the distance from device $i$ to the \gls{bs} in km, and $\Upsilon_{i}$ represents the random variations in \gls{lsfc}, e.g., shadow fading, with distribution $\cN(0,\sigma_{\textup{sf}}^2)$ -- this is the 3GPP Urban Microcell model in \cite{3gpp.36.814} with a carrier frequency of 2~GHz, and we set $\sigma_{\textup{sf}}^2=8$	dB. When generating the \glspl{lsfc}, we use a wrap-around technique by drawing 6 cells around the central cell and setting the \gls{lsfc} of a user to be the largest one among the \glspl{lsfc} to all the \glspl{bs}. The maximum transmit power is $\rho_{\max}=23$~dBm \cite{3gpp.36.331}. The noise spectral density is $-169$~dBm/Hz and the system bandwidth is $180$~kHz \cite{3gpp.36.331}. The rate penalty factor is set to $\ell=0.25$ to give a close approximation to the normal approximation in \cite[Th. 55]{polyanskiy2010channel}.

For the log-sum-exp approximation in \eqref{S1}, we set $\alpha=3$ for the fairness promoting function in \eqref{eq: fairness func}, and the frame length is set to $T=20$. For the virtual-queue-based approximation in \eqref{S2}, we set $V=1000$ and $z_{\max}=100$ in \eqref{eq: z}. The size of all hidden layers in the neural network is set to $64$. The training is performed using RMSprop with learning rate $5\times10^{-4}$, with smoothing constant $0.99$, and without weight decay or momentum. To avoid exploding gradients, we perform gradient clipping on the GRU layer and set the maximum gradient norm to $10$. 

We run $1000$ training epochs, each has $100$ episodes with $T=20$ slots. The generated episodes are stored in a replay buffer of size $5000$. Each training epoch is followed by $100$ training steps performed on a mini-batch of $|\cS|=32$ episodes randomly sampled from the replay buffer. After training, we run $200$ testing epochs.

We consider the following two performance metrics:
\begin{itemize}
    \item \emph{Maximum NCPDR}: The objective in the original problem \eqref{P}, given by $\max_{i\in\cN}\{{\overline{D}_i}/{D_i^\textup{th}} \}$. It characterizes the performance of the worst-performing device (fairness). 
    \item \emph{Sum effective throughput}: the sum of the effective throughput of all devices, i.e., $\sum_{i\in\cN}(\lambda_i - {\overline{D}_i})$. It characterizes the overall performance of the network.
\end{itemize}

\subsection{Performance Evaluation}

\begin{figure*}
	\centering
	\includegraphics[]{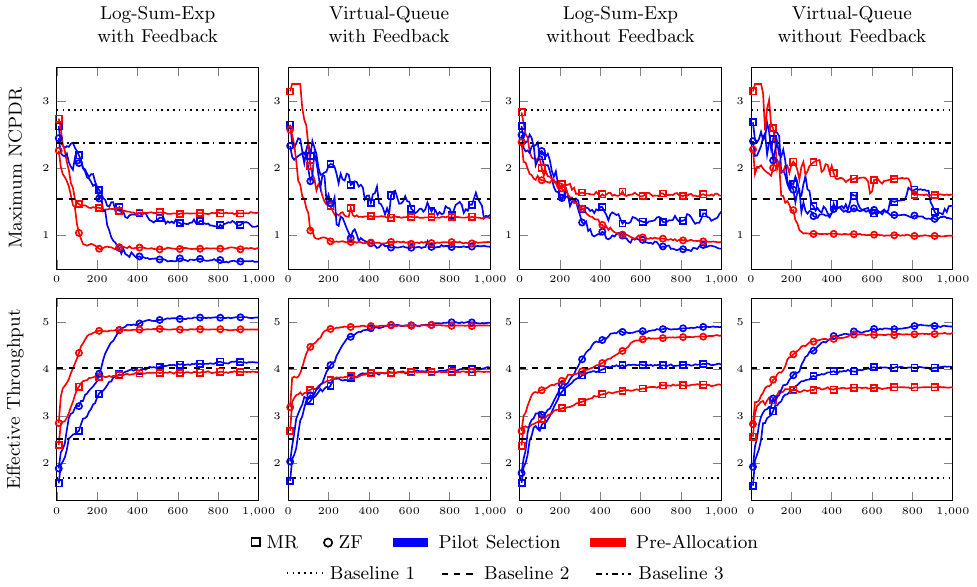}
	\caption{Performance comparison (averaged over 8 independent trials and over every 10 epochs).}
	\label{result: performance}
\end{figure*}

We first consider a system with $N=12$ devices and $L=6$ pilots. 
The devices are divided into two classes based on their \emph{heterogeneous} traffic and QoS requirements:
\begin{itemize}
    \item \emph{Class 1}: For $i\in\{1,2,3,4\}$, the packet arrival rate is $\lambda_i=0.2$ packets/slot. The packet drop rate threshold is $D_i^\textup{th}=0.05$ packets/slot. The data rate requirement is $R_i^\textup{th}=1$ bits/s/Hz. Each packet expires in $d_i^{\max}=2$ slots.
    \item \emph{Class 2}: For $i\in\{5,\cdots,12\}$, the packet arrival rate is $\lambda_i=0.65$ packets/slot. The packet drop rate threshold is $D_i^\textup{th}=0.2$ packets/slot. The data rate requirement is $R_i^\textup{th}=2$ bits/s/Hz. Each packet expires in $d_i^{\max}\!=\!5$ slots.
\end{itemize}

To benchmark the performance, we consider the following three baseline approaches:
\begin{itemize}
	\item \emph{Baseline 1}: We assume that a genie knows the number of backlogged users $|\cK_t|$. It informs each backlogged user the optimal access barring parameter $p_{\textup{bar}} = \min\{L/{|\cK_t|}, 1\}$. At the beginning of a slot, each backlogged user generates a random number $p$ uniformly in $[0,1]$. The user transmits a randomly selected pilot if $p<p_{\textup{bar}}$ and transmits the head-of-line packet using full power (we observed better performance than using the channel inversion power control). The \gls{bs} performs \gls{zf} combining.
	\item \emph{Baseline 2}: We assume scheduled transmissions to avoid collisions. Specifically, we pre-allocate the same pilot to user $i$ and user $i+6$, for $i\in\{1,\cdots,6\}$. The users that share the same pilot will transmit in turn -- users $i\in\{1,\cdots,6\}$ can transmit in even slots and the other users transmit in odd slots if they are backlogged. The active users use full power to transmit their payload data. The \gls{bs} performs \gls{zf} combining.
    \item \emph{Baseline 3}: 
    Instead of reusing the mutually orthogonal pilots, another scheme is to use pre-assigned, unique but \emph{non-orthogonal} pilots. 
    Specifically, each user $i\in\cN$ is assigned a pilot sequence $\bpsi_i$ of unit energy. 
    Since state-of-the-art activity detection algorithms for non-orthogonal pilots have shown remarkable performance \cite{fengler2022pilot}, we assume that all active users can be correctly detected. 
    Denoting by $\bPsi \defeq [\bpsi_1,\cdots,\bpsi_N]$ and $\widetilde{\bPsi}_t \defeq \sqrt{\rho_0}(\bPsi_t^\act)^*$,\footnote{Analogous to Definition \ref{def: act}, the superscript $(\cdot)^\act$ is used to represent the elements corresponding to active users $\overline{\cU}_t$.} the \gls{mmse} estimate of the user channel matrix $\bH_t \defeq [\bh_{1t},\cdots,\bh_{Nt}] \in \C^{M\times N}$ is
    \begin{equation}
    	\widehat{\bH}_t^\act = \bY_t^\p\widetilde{\bPsi} \left(\widetilde{\bPsi}^\herm\widetilde{\bPsi} + \bI \right)^{-1},
    \end{equation}
 	where $\bY_t^\p = [\by_{1t},\cdots,\by_{Mt}]^\transp$.  Notice that, unlike the case of orthogonal pilots in \eqref{eq: channel estimate orthogonal}, the channel estimates do not decouple across users and become linearly dependent.
    We can perform \gls{mr} and \gls{zf} combining by using the combining matrix
    \begin{equation}
		\bV_t^\act \defeq \left\{
    	\begin{array}{ll}
       		\widehat{\bH}_t^\act, & \text{MR} \\
        	\widehat{\bH}_t^\act \left((\widehat{\bH}_t^\act)^\herm\widehat{\bH}_t^\act \right)^{-1} , & \text{ZF}
    	\end{array}
    	\right..
	\end{equation}
    We apply the \gls{zf} combining by default. However, the \gls{zf} combining does not work when $|\overline{\cU}_t| > L$, since  the columns of $\widehat{\bH}_t^\act$ become linearly dependent so that $(\widehat{\bH}_t^\act)^\herm\widehat{\bH}_t^\act$ becomes singular. In this case, we can only use \gls{mr} combining. We use the same access barring scheme as in Baseline 1.

\end{itemize}

\begin{figure*}
    \centering
    \includegraphics[width=16.5cm]{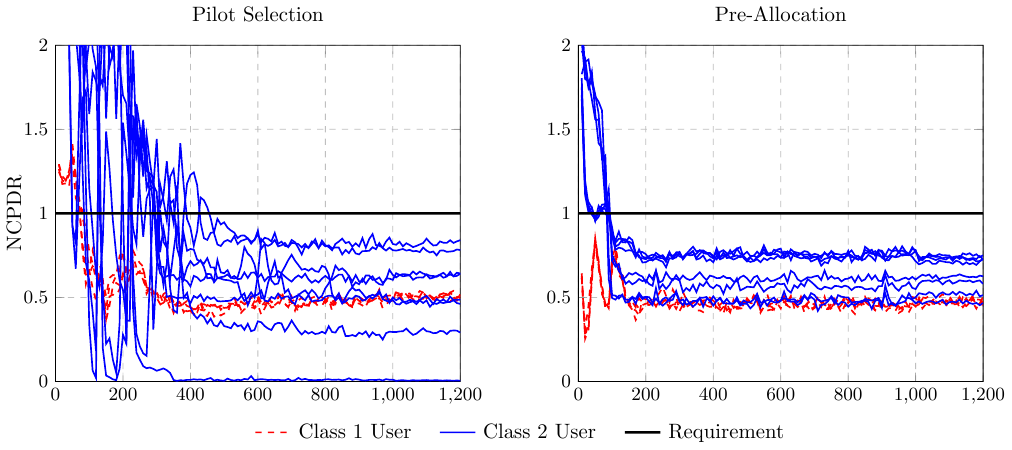}
    \caption{Normalized packet drop rate per user (single trial, averaged over 10 epochs). The requirement line represents $\overline{D}_i/D_i^{\textup{th}} = 1$.}
    \label{fig: dr_per_user}
\end{figure*}

We consider the same feedback message as in \cite{9204466}, which contains a ternary indicator (successful transmission, collision, and idle) for each pilot.
We also consider the pilot pre-allocation with the same allocation pattern as in Baseline 2 but without scheduling. The pilot selection reduces to on-off decisions when using pre-allocation.
The performance achieved by different schemes during different training epochs is summarized in Fig. \ref{result: performance}.   
We make the following observations. 
Both the log-sum-exp and the virtual-queue approximations can provide fairness among users, while the former works slightly better. 
Using pilot status as feedback information accelerates the convergence and improves the final performance. 
Pilot pre-allocation accelerates the training with a slight performance loss. 
Compared with MR, ZF combing achieves significantly better performance by reducing the interference power, and the loss of spatial degrees of freedom is negligible due to the large number of antennas. 
In Fig. \ref{fig: dr_per_user}, we plot the packet drop rates of each user during training with or without pilot pre-allocation, using the log-sum-exp approximation and  ZF combining.

\begin{figure*}
     \centering
     \begin{subfigure}[b]{7cm}
         \includegraphics[width=\textwidth]{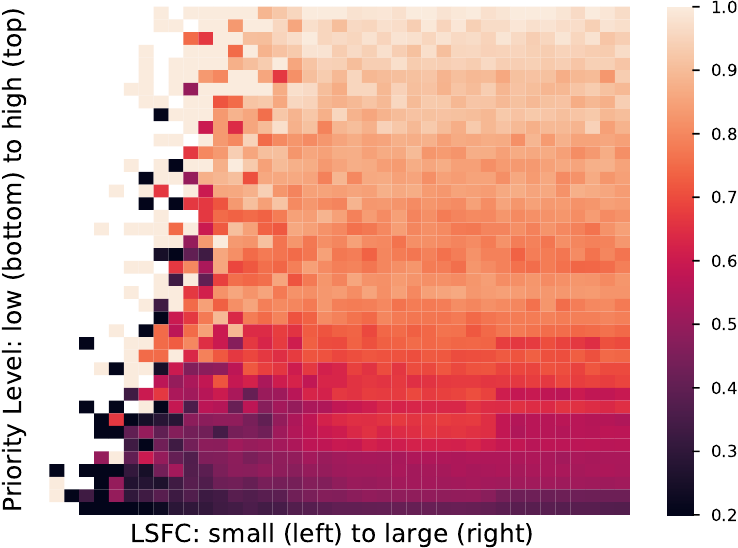}
         \caption{pilot access}
         \label{fig: pilot}
     \end{subfigure}
     \hspace{1cm}
     \begin{subfigure}[b]{7cm}
         \includegraphics[width=\textwidth]{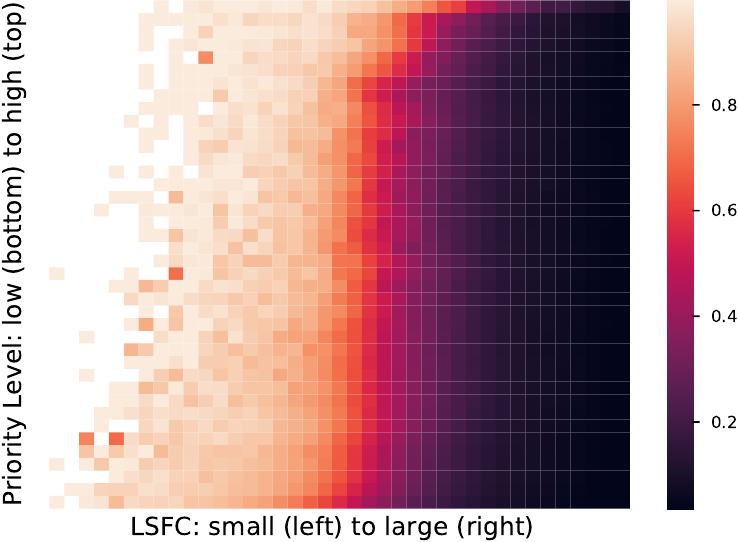}
         \caption{power control}
         \label{fig: power}
     \end{subfigure}
     \caption{Visualization of the learned policy.}
     \label{fig: policy}
\end{figure*}

The learned policies in Fig. \ref{fig: dr_per_user} with pilot pre-allocation are visualized in Fig. \ref{fig: policy}. Our purpose is to see how the user status will affect the policy outputs (the access probability and the transmit power). To do this, we collect all the policy outputs during the testing epochs and calculate the average values for each given priority level and LSFC (which are uniformly quantized in dB) and plot them as heat maps. As shown in Fig. \ref{fig: pilot}, a user has higher access probability when its priority level is high, and becomes more conservative for low priority levels. The LSFC also has impact on the access probability. In Fig. \ref{fig: power}, we can observe that users use larger transmit power when the LSFC is small (consistent with most of power control schemes), and extreme priority levels will also affect the transmit power. Notice that this visualization shows only the impact on average. The learned policy could be much more complicated due to the temporal correlation.

\subsection{Does our learning framework scale?}

\begin{figure}[!t]
    \centering
      \centering
      \includegraphics[width=8cm]{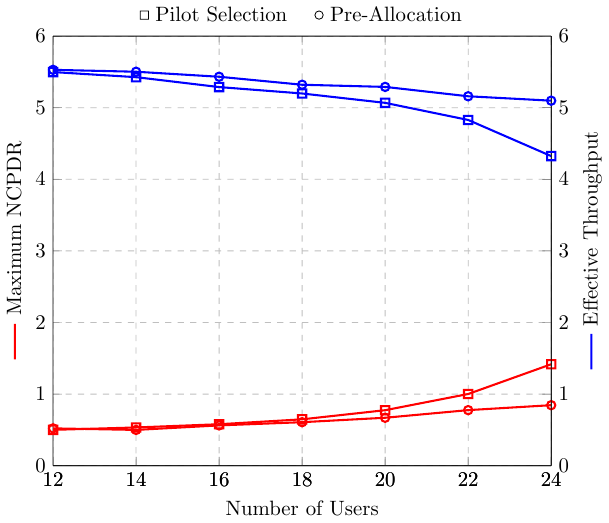}
      \caption{Scalability results.}
      \label{fig: scalability}
  \end{figure}
  
  \begin{figure*}
      \centering
      \includegraphics[]{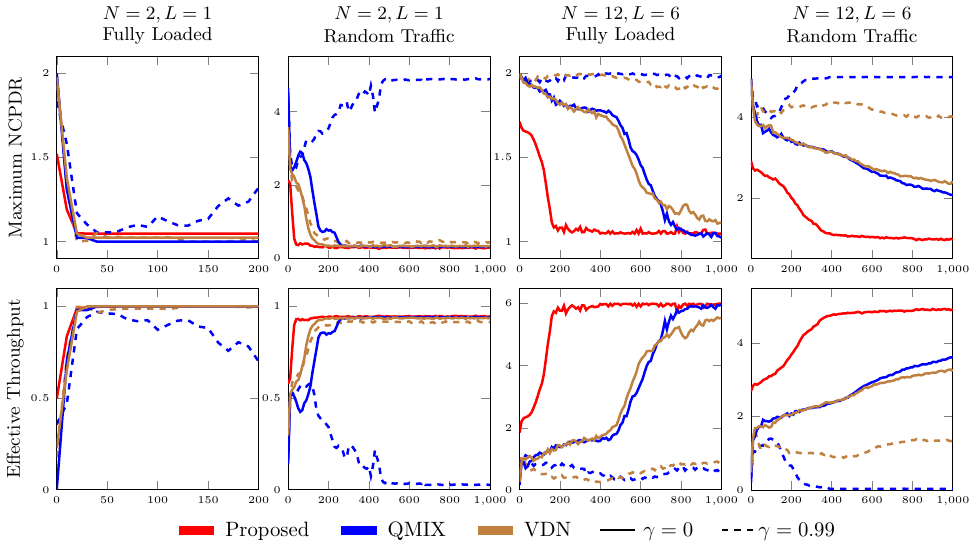}
      \caption{Comparison with VDN and QMIX. (Averaged over 4 independent trials and over every 10 epochs.)}
      \label{fig: comparison}
  \end{figure*}
  
Scalability is always a critical aspect of multi-agent learning frameworks. When complicated competition and cooperation exist among agents, the frameworks usually do not scale well. The pilot collision represents a very strong interaction, and it is difficult to train for a system with hundreds or thousands of users. Our framework, although more efficient than conventional RL in our particular scenario, also suffers from performance loss due to the limited scalability. One remedy is to limit the interactions among agents. As a showcase, we consider a system with $L=6$ pilots, and the number of users, $N$, varies from $12$ to $24$. The packet arrival rate is $L/N$ packet/slot, the drop rate threshold is $1.2/N$ packets/slot, and the rate requirement is $1.5$ bit/s/Hz, for all users. We consider two schemes: 1) each user can select any of the pilots, and 2) the users are divided into two groups each with half of the users, and each group is pre-allocated $3$ pilots. We fix the number of training epochs to $1000$ and evaluate the performance by averaging over $200$ testing epochs. We use the log-sum-exp approximation, ZF processing, and the feedback message. The results are shown in Fig. \ref{fig: scalability}. We observe that, by limiting the number of training resources, the second scheme scales better. Our learning framework is more suitable for a small number of high-priority users with stringent performance requirements, while other solutions (e.g., cluster-based scheduling) and more scalable approaches are necessary for large-scale systems.

\subsection{Comparison with RL} \label{sec: comparison with RL}
We compare the proposed learning scheme with VDN \cite{VDN} and QMIX \cite{QMIX}, two standard benchmarks for cooperative \gls{marl} with team reward. 
Since VDN and QMIX do not natively support hybrid policies, we ignore the data transmission part and consider a collision model -- the transmission is successful as long as the selected pilot is not occupied by other users.
We consider two different system sizes, $(N=2,L=1)$ and $(N=12,L=6)$, and two traffic models:
\begin{itemize}
    \item \emph{Fully-loaded}: Each user generates a packet in each slot, i.e., $\lambda_i=1$ packet/slot for all $i\in\cN$. The packet drop rate threshold is $D_i^\textup{th}=0.5$ packets/slot. Each packet expires immediately after the current slot, i.e., $d_i^{\max} = 1$. It is a simple case, where the users always have packets and they only need to learn to cooperate in a static environment to avoid collisions and achieve fairness by giving up half of the transmission opportunities.
    \item \emph{Random Traffic}: Each user randomly generates a packet with probability $\lambda_i = 0.5$ in each slot. The packet drop rate threshold is $D_i^\textup{th}=0.1$ packets/slot. Each packet expires in $d_i^{\max}=5$ slots. Compared to the fully-loaded system, the users also need to learn to predict and adapt to the environment changes (the queue status) and satisfy the delay constraints.
\end{itemize}

We implement VDN and QMIX based on the code available at \url{https://github.com/oxwhirl/pymarl}.
The only difference in the network structure is that we replace the output layer in the agent network, which is a single feedforward layer with linear activation in the original implementation, by two feedforward layers with a ReLU activation function for the first layer.
The other learning parameters are set to be the same as with the proposed approach, which also match the default settings in the original implementation.
After all active users select their transmission actions, we use the obtained objective value in \eqref{S1} as the team reward for VDN and QMIX.
The parameter of the log-sum-exp approximation is set to $\alpha=15$ in the fully-loaded system and $\alpha=3$ for random traffic.
 
To investigate the trade-off between long-term planning and the adopted greedy scheme for this problem, we consider two discount factors, $\gamma=0$ and $\gamma=0.99$, for VDN and QMIX.
When $\gamma=0$, the agents only need to estimate the expected immediate reward function and select the actions to greedily maximize it when making a decision. When $\gamma=0.99$, the agents consider the long-term return, and they need to estimate the discounted sum of future rewards, which requires more exploration.
For exploration in VDN and QMIX, we adopt an $\epsilon$-greedy policy, where the users select random actions with probability $\epsilon$ when generating episodes for training. (During testing after each epoch, the users always select the action with the highest estimated value.) 
Similar to \cite{QMIX}, we anneal $\epsilon$ linearly from 1 to 0.05 during training.
Based on the scenario, we set the annealing time to 10 epochs for $(N=2,L=1)$ in the fully-loaded system, 100 epochs for $(N=2,L=1)$ with random traffic, and 500 epochs when $(N=12,L=6)$.

The performance comparison is shown in Fig. \ref{fig: comparison}.
When the system is small (first two columns in Fig. \ref{fig: comparison}), all algorithms (except QMIX with $\gamma=0.99$) can efficiently learn a cooperative policy to avoid collisions and achieve good fairness between the two users.
When the system becomes larger but remains relatively static (third column in Fig. \ref{fig: comparison}), VDN and QMIX can still learn a cooperative policy with $\gamma=0$, but the proposed approach can learn much more efficiently. However, VDN and QMIX with $\gamma=0.99$ fail to learn a useful policy.
In the most challenging scenario where the system is large and has highly dynamic traffic (last column in Fig. \ref{fig: comparison}), the proposed approach can still learn efficiently, while VDN and QMIX struggle. When $\gamma=0$, the performance of VDN and QMIX still slowly improves after 1000 training epochs, but it may take much longer to converge.

There are some interesting observations from the comparison that we would like to highlight:

\subsubsection{Long-term planning v.s. greedy scheme}
In our development of \eqref{S1} and \eqref{S2}, we choose to greedily maximize the immediate objective function when making each decision. 
Long-term planning is usually preferred in RL, as greedily maximizing the immediate reward may prevent the agents to select better actions in the future.
However, the design of our objective function is quite different from the conventional RL reward function -- we have already incorporated the urgency level of packets, which is the most critical factor for future planning. 
For our particular problem, we do not see what other factors may have significant effects in the long run, as future packet arrivals are independent of the current state and decisions.
Sending packets that are most urgent while prioritizing fairness also does not seem to prevent the users from selecting better alternatives in the future.
In this sense, our design of the objective function is more analogous to the value function in RL instead of the immediate reward function, and there is no need for additional long-term planning when implementing the \gls{rl} algorithms.
In the simulation results in Fig. \ref{fig: comparison}, we also observe that choosing the greedy scheme ($\gamma=0$) works better than long-term planning ($\gamma=0.99$) in all considered scenarios.

\subsubsection{Exploration v.s. guided learning}
In conventional RL, exploration is essential to find good actions to be reinforced.
Specifically, the agents need to take random actions to obtain a good estimate of the value function at the beginning of the training.
In small-scale systems, the chance to randomly take a good joint action is high, and the exploration can be effective.
However, as the system becomes larger, the exploration becomes more challenging, especially when the system is dynamic.
In contrast, our model-based approach is more efficient due to the closed-form, differentiable training objective, which enables us to directly optimize the policy without the need for trying random actions.
The effectiveness of the proposed approach against conventional RL is verified in the simulation results in Fig. \ref{fig: comparison}.
Our approach also seamlessly integrates discrete pilot selection decisions and continuous power control with data rate requirements, which, to the best of our knowledge, has not been done before.

\begin{remark}
    We consistently observe that the training of QMIX with $\gamma=0.99$ is unstable and does not converge in our simulations.
    Even in the simplest case for $(N=2,L=1)$ with fully-loaded traffic, it first finds a good policy but then diverges as the training progresses.
    We have tried different learning rates (from $10^{-3}$ to $10^{-5}$) and different structures of the mixing network, but the problem persists.
    We suspect that this is due to the unnecessity of extra long-term planning in our problem, as discussed above, and because the additional expressibility of the mixing network may result in a compromised factorization of the joint value function.
    As the chosen reward function (objectives in \eqref{S1}) is already in the form of a sum of individual contributions, it is more suitable for VDN, where the factorization is forced to be a sum.
\end{remark}

\begin{remark}
    Another approach that considers only the immediate reward for a given situation is \gls{cbl}, which is a special case of full RL \cite[Ch. 2]{sutton2018reinforcement}. 
	In CBL, taking an action will only affect the immediate reward, instead of future states as in full RL, and the agents share the same observation of the context.
    This is conceptually different from our considered scenario, where the transmission decision will affect the next state (i.e., the queue backlogs and the urgency levels of the remaining packets), and the users do not share the same observation.
\end{remark}

\section{Conclusion}
\label{sec: conclusion}
In this work, we provide a cross-layer \gls{gfra} model with MIMO and dynamic traffic. 
We formulate a fairness-based stochastic network optimization problem and develop two real-time approximations to this stochastic problem. 
These approximations give a unified measure of instantaneous fairness among users. 
We develop a distributed policy that seamlessly combines discrete pilot selection decisions and continuous power control variables to maximize user fairness and network performance. 
In contrast to conventional sample/exploration-based RL approaches, our training objective (expected reward) is differentiable with respect to the policy parameters and thus allows more efficient training. 
Our work suggests that one can achieve considerable performance improvements by incorporating domain knowledge and model structure into the learning design.

\appendix

Since we consider only a single slot here, we omit the time indices for brevity. When using ZF, for a non-collided user $i$, we have $\bv_i^\herm\widehat{\bg}_{a_i} = 1$, and $\bv_i^\herm\widehat{\bg}_{a_j}=0$ when $j\neq i$. This gives
\begin{equation}
\begin{aligned}
\label{eq: expected inverse SINR for ZF 1}
	\expt{}{\frac{1}{\SINR_i}} = \E\Bigg[&\expt{}{|\bv_i^\herm\widetilde{\bg}_{a_i}|^2\Big|\widehat{\bG}}  + \frac{1}{\beta_i\rho_i}\|\bv_i\|^2 \\
	&+ \sum_{j\in\activeusers\backslash i}\frac{\beta_j\rho_j}{\beta_i\rho_i}\expt{}{|\bv_i^\herm\bh_j|^2\Big|\widehat{\bG}}\Bigg].
\end{aligned}
\end{equation}
Notice that $\bv_i$ becomes a constant vector when conditioned on $\widehat{\bG}$. To evaluate the first conditional expectation, we note that $\widetilde{\bg}_{a_i}$ is always independent of $\widehat{\bG}$ regardless of the employed pilots, and therefore, 
$\bv_i^\herm\widetilde{\bg}_{a_i} \sim \cn{0}{(1-c_{a_i})\|\bv_i \|^2}$; hence
\begin{equation}
\label{eq: ZF term1}
	\expt{}{|\bv_i^\herm\widetilde{\bg}_{a_i}|^2\Big|\widehat{\bG}} = (1-c_{a_i})\|\bv_i \|^2.
\end{equation}
By using \eqref{eq: channel estimate orthogonal}, we have 
\begin{equation}
\begin{aligned}
	\widehat{\bg}_{a_j} =& \frac{\rho_0|\cU_{a_j}|}{1 + \rho_0|\cU_{a_j}|}\frac{1}{\sqrt{|\cU_{a_j}|}}\sum_{k\in\cU_{a_j}} \bh_k + \frac{\sqrt{\rho_0|\cU_{a_j}|}}{1 + \rho_0|\cU_{a_j}|} \bw\\
	=& \frac{c_{a_j}}{\sqrt{|\cU_{a_j}|}}\sum_{k\in\cU_{a_j}} \bh_k + \sqrt{c_{a_j}(1-c_{a_j}) } \bw,
\end{aligned}
\end{equation}
where $\bw\sim\cn{\bzero}{\bI}$ and $\{\bh_j\}$ are mutually independent. This tells that $\expt{}{\bh_j\widehat{\bg}_{a_j}^\herm}=\frac{c_{a_j}}{\sqrt{|\cU_{a_j}|}}\bI$. 
Since $\bh_j$ and $\widehat{\bg}_{a_j}$ are jointly Gaussian, we know from \cite[Theorem 10.2]{kay1993fundamentals} that $\frac{1}{\sqrt{|\cU_{a_j}|}}\widehat{\bg}_{a_j}$ is the \gls{mmse} estimate of $\bh_j$ given $\widehat{\bg}_{a_j}$. By the orthogonality principle, we can write 
$
	\bh_j = \frac{1}{\sqrt{|\cU_{a_j}|}}\widehat{\bg}_{a_j} + \bz_j,
$
where $\bz_j$ has distribution $\cn{\bzero}{\left(1-\frac{c_{a_j}}{|\cU_{a_j}|}\right)\bI}$ and is independent of $\widehat{\bG}$. 
The second conditional expectation is then evaluated as 
\begin{equation}
\label{eq: ZF term2 orthogonal}
	\expt{}{|\bv_i^\herm\bh_j|^2\Big|\widehat{\bG}} = \left(1-\frac{c_{a_j}}{|\cU_{a_j}|} \right)\|\bv_i \|^2.
\end{equation}
By substituting \eqref{eq: ZF term1} and \eqref{eq: ZF term2 orthogonal} into \eqref{eq: expected inverse SINR for ZF 1}, we obtain
\begin{equation}
\begin{aligned}
\label{eq: expected inverse SINR for ZF 2}
	\expt{}{\frac{1}{\SINR_i}} = &\frac{\expt{}{\|\bv_i\|^2}}{\beta_i\rho_i}\Bigg((1-c_{a_i})\beta_i\rho_i  + 1\\ 
	&+ \sum_{j\in\activeusers\backslash i}\left(1-\frac{c_{a_j}}{|\cU_{a_j}|} \right)\beta_j\rho_j \Bigg).
\end{aligned}
\end{equation}

The final step is to evaluate $\expt{}{\|\bv_i \|^2}$, which is given by
\begin{equation*}
	\expt{}{\|\bv_i \|^2} = c_{a_i}^{-1}\left[\expt{}{(\bQ^\herm\bQ)^{-1}}\right]_{i,i} = \frac{1}{c_{a_i}(M-|\cL^\act|)}
\end{equation*}
where $\bQ$ is a $M\times|\cL^\act|$ matrix with independent $\cn{0}{1}$ entries, and the second equality follows immediately from \cite[Appendix B]{marzetta2016fundamentals}. 
\hfill $\IEEEQED$

\bibliographystyle{IEEEtran}
\bibliography{ref}

\end{document}